\documentclass{article}
\usepackage{graphicx}
\usepackage{anyfontsize}
\graphicspath{ {./draft_jpeg/} }
\begin{document}

\newcommand{\D}{\ensuremath\mathrm{d}}
\begin{center}
{\large \bf Deformation-induced charge redistribution in Ceria thin film at room temperature}

{Kyoung-Won Park$^{a,*}$, Chang Sub Kim$^{b}$}

{\small $^{a}$ Center for Biomaterials, Korea Institute of Science and Technology, Seoul, 02792, Republic of Korea}\\
{\small $^{b}$ Department of Materials Science and Engineering, Massachusetts Institute of Technology, Cambridge, Massachusetts 02139, USA}\\

\end{center}

\begin{abstract}

Tuning electronic properties through strain engineering of metal oxides is an important step toward understanding electrochemical and catalytic reactions in energy storage and conversion devices. Traditionally, strain engineering studies focused on movement of oxygen ions at high temperatures (500 degree Celcius and above), complicating electrical properties by introducing mixed electronic and ionic conductivity. In this study, we demonstrate room temperature charge redistribution in a CeO$_{2}$ thin film as a result of phase transformation in a localized region by mechanical deformation. Mechanical indentation of the CeO$_{2}$ thin film at room temperature results in irreversible deformation. Conductive-tip atomic force microscopy (C-AFM) analysis shows increased current passing through the locally deformed area of the CeO$_{2}$ thin film. Electron energy loss spectroscopy (EELS) analysis equipped with transmission electron microscopy (TEM) suggests that the increase in current contrast in the deformed region arises from an increased concentration of Ce$^{3+}$ ions. We herein discuss the fundamental reason behind the increased amount of Ce$^{3+}$ ions in the deformed area, based on the atomic scale computational works performed by molecular dynamics (MD) simulations and first-principles density functional theory (DFT) calculations. Plastic deformation induces a phase transformation of cubic fluorite CeO$_{2}$ into a newly formed T-CeO$_{2}$ structure. This phase transformation occurs mainly by oxygen ions moving closer to cerium ions to release mechanical energy absorbed in the CeO$_{2}$ thin film, followed by charge redistribution from the initial CeO$_{2}$ to the newly created T-CeO$_{2}$ structure.\\
\end{abstract}

{\textbf{keywords}}: CeO$_{2}$ thin film; mechanical deformation; phase transformation; oxidation state; electronic characteristics; charge redistribution\\

\vfill

Corresponding Author\\
$^{*}$E-mail: exclaim27@kist.re.kr\\

\newpage

\section{Introduction}
The fact that electronic characteristics of space charge layers differ from bulk phases [1,2] is considered the key factor in tuning various properties of photocatalysts [3], gas sensors [4-6], batteries [7-9], and solid-state fuel cells (or ion conductors) [10,11]. For example, gas sensors generally sense current change, the magnitude of which varies depending on the extent of band bending caused by formation of the space charge layer [12,13] during electrochemical equilibration between the adsorbate and the sensing surface. Depending on the surface chemistry and material in photocatalytic water splitting, different space charge layers form in an aqueous solution, resulting in different aspects of band bending at the surface/water interface [14,15]. Different band bending also varies the band edge positions relative to water redox potentials at the interface, which leads to a variety of light absorption and photocatalytic behaviors [16,17]. The aspect of the space charge layer (or band bending) critically determines charge separation in the photocatalyst, directly affecting the photocatalytic efficiency [18-20]. Structural deviation from bulk regions in electrode materials, including grain boundaries [10,21] and heterointerfaces [22-24], also forms the space charge layer, resulting in varied electronic conductivity.

Strain engineering, including lattice mismatch and various dimensional (1-D and 2-D) defect formations (e.g., dislocations and grain boundaries) has emerged as an interesting approach to tune the electronic characteristics of metal oxides used in electrochemical catalysts [25,26], solid oxide fuel cell (SOFC) electrodes [27-31], solar fuel generation [32-34], and batteries [35,36]. However, many studies have focused on the oxygen vacancy-mediated space charge layer, whether strain engineering improves or retards oxygen mobility in the lattice [28,29,31,37,38], and how modulated oxygen mobility is related to formation of the space charge layer. Therefore, diffusion behavior of oxygen ions or oxygen ion conductivity has been examined at elevated temperature to detect the modulation due to the electrically insulating nature of a variety of metal oxide materials at room temperature. 

However, it is difficult to find the origin of varied electrical conductivity induced by strain engineering at high temperature because electrical conductivity at elevated temperature can be affected by a variety of factors, such as the formation of structural defects (e.g., oxygen vacancies, dislocation, grain boundaries, and twinning), the concentration variation of oxygen vacancies, the redistribution of oxygen vacancies around various dimensional defects, and change in electron hopping behavior due to interactions between oxygen vacancies and defects. In fact, it is still arguable whether electrical conductivity indeed increases (or decreases) with strain engineering, both from oxygen transport [24,37,39] and electron hopping [40] points of view. Thus, it is unknown whether variation in electrical conductivities induced by strain engineering at elevated temperature comes from oxygen ion diffusion, inhomogeneous formation of structural defects, change in electronic conductivity originating from redistribution of oxygen vacancies, or a combination of them.
To study the sole (intrinsic) effect of strain on the electronic characteristics of metal oxides, we herein investigate the change in the electrical current after local deformation of CeO$_{2}$ thin films by indentation at room temperature, under the assumption that oxygen vacancy concentration is very low and oxygen ion diffusion is negligible (i.e., by eliminating the effect of oxygen vacancy redistribution). Conductive-tip atomic force microscopy (C-AFM) analysis of the indented CeO$_{2}$ thin film at room temperature indicates that electrical current passing through the thin film is locally enhanced near the indented area. The change in current is demonstrated to come from the increased Ce$^{3+}$/Ce$^{4+}$ ratio in the deformed region according to transmission electron microscopy-electron energy loss spectroscopy (TEM-EELS) analysis, suggesting that local deformation induces modulation of the oxidation state of the Ce cation. Based on computational evidence from first-principles density functional theory (DFT) calculations and molecular dynamics (MD) simulations of CeO$_{2}$, we discuss in depth the fundamental reason for the enhanced current as well as the Ce$^{3+}$/Ce$^{4+}$ ratio observed experimentally from deformation, taking atomic structural change and charge redistribution points of view. 

Based on our fundamental study, we first suggest that structural changes induced by mechanical deformation at room temperature can cause heterogeneous charge distribution, resulting in oxidation state change of the cation as well as electronic conductivity, which differs from the long-standing mechanism of oxygen vacancy-mediated transport at elevated temperature. We expect that this mechanical method can be employed to easily prepare heterogeneous structures that have different phases and electronic characteristics; the same effect as introducing non-stoichiometric grain boundaries in a homogeneous electrode material can be obtained. In addition, optimal band edges of photocatalysts such as tandem cells can be achieved in a simple mechanical manner so that the heterogeneous photocatalysts can work best to achieve higher energy conversion efficiency.

\section{Methods}

\subsection{Model system}
Fluorite structured CeO$_{2}$ was selected as a model system because this system easily changes oxidation state of the Ce ion between Ce$^{4+}$ and Ce$^{3+}$ by oxygen vacancy formation/annihilation under environmental changes (e.g., doping [41-43], oxygen partial pressure [44], temperature change [44,45], or surface generation [46]). This facile redox ability of the Ce ion is suitable for detecting the change in electronic characteristics (or space charge layer formation). It is also an electrical insulator at room temperature. Hence, we can assume that oxygen vacancy concentration and diffusivity at room temperature are low enough to ignore, resulting in a clear analysis of the predominant effect of mechanical deformation on the electronic properties at room temperature.

\subsection{Thin film deposition}
A thin layer of Pt (200 nm thick), serving as a current collector, was deposited on a (001) oriented single crystal of 8 mol$\%$ Y$_{2}$O$_{3}$ stabilized ZrO$_{2}$ (YSZ) substrates ($10 \times 5 \times 0.5$ mm$^{3}$; MTI Corporation, Richmond, CA) by DC magnetron sputtering (Kurt J. Lesker, Clairton, PA) in a sputtering chamber evacuated to a background pressure of $<$ $5 \times 10$$^{-}$$^{6}$ Torr using a cryogenic pump (CTI Cryogenics, Cryotorr 8, Chelmsford, MA). A Pt sputter target (purity $>$ 99.99 wt$\%$, Kurt J. Lesker) was used for deposition after surface cleaning with a plasma power of 50 W for 3 min. Ar at 10 mTorr was maintained during deposition using mass controllers (MKS Instruments, 1179A, Wilmington, MA) and a process controller (MKS Instruments, 647C) with a plasma power of 50 W at room temperature.

A thin film of CeO$_{2}$ (120 nm thick) was deposited by pulsed laser deposition (PLD) onto the Pt decorated YSZ substrate (Pt/YSZ). The PLD system (Neocera Inc., Beltsville, ML) was operated with a KrF excimer laser (Coherent COMPex Pro 205) emitting at 248 nm with an energy of 400 mJ/pulse and with a repetition rate of 10 Hz. The Pt/YSZ substrates were heated to 650 degree Celcius while the oxygen pressure was maintained at 10 mTorr during deposition after pumping the background pressure to $<$ $9 \times 10$$^{-}$$^{6}$ Torr. Prior to cooling, the oxygen pressure in the chamber was increased to approximately 10 Torr to facilitate more complete oxidation of the films. The thickness of the CeO$_{2}$ thin film deposited on the Pt/YSZ was evaluated using a profilometer (Bruker DXT Stylus Profilometer). The prepared thin film/substrate (CeO$_{2}$/Pt/YSZ ; $10 \times 5 \times 0.5$ mm$^{3}$) was divided into $\approx$ 9 pieces ($5 \times 1 \times 0.5$ mm$^{3}$) using a diamond cutter (DISCO ADA-321 dicing saw) to prepare small samples to more easily identify indents in imaging/milling with dual beam focused ion beam-scanning electron microscopy (FIB-SEM) and C-AFM analysis after indentation.


\subsection{Local deformation: Indentation}
On prepared CeO$_{2}$/Pt/YSZ, indentation experiments were performed at room temperature using a Tribo-indenter (Hysitron Inc., Minneapolis, MN) with a Berkovich tip (150 nm average radius of curvature). Load was applied on CeO$_{2}$/Pt/YSZ at a rate of 200 mm/s up to a peak load ($P_{max}$) of 9000 $\mu$N. The indentation was repeated 1350 times by maintaining the distance between adjacent indents at 10 $\mu$m. In the loading-unloading process, the load versus displacement was recorded, as shown in Figure S1a.

\subsection{Current change analysis: CT-AFM}
Indented CeO$_{2}$/Pt/YSZ was analyzed with C-AFM to detect the current change in the local area using a commercial AFM system (Asylum Cypher ES) and a conductive diamond-coated silicon tip (Nanosensors CDT-FMR, resonance frequency of $\approx$120 kHz) that has good conductivity and is resistant to wear. To electrically bias a sample, the indented CeO$_{2}$/Pt/YSZ was attached to an electrical sample puck assembly (part number: 448.140) using silver (Ag) paste. In the Ag pasting process, the Pt current collecting layer in CeO$_{2}$/Pt/YSZ was connected to the electrical sample puck assembly (Figure S2). For scanning current passing through from the tip to the sample (i.e., for C-AFM imaging), an ORCA cantilever holder was employed in basic AC and contact mode (ORCA mode) with at least 1 pA current sensitivity. To clarify the origin of the current contrast from CeO$_{2}$/Pt/YSZ, C-AFM analysis was also carried out on Pt/YSZ indented under the same conditions (Figure S1b).

\subsection{Oxidation state of Ce ion: TEM-EELS analysis}
A TEM sample of the indented CeO$_{2}$/Pt/YSZ was prepared by using a focused ion beam (FIB, FEI Helios 600 NanoLab). Prior to placing the indented CeO$_{2}$/Pt/YSZ in the vacuum chamber of FIB-SEM, Pt/Pd (80:20; 30 nm) was deposited on indented CeO$_{2}$/Pt/YSZ using a sputter coater (EMS 300TD) to eliminate charge effects and to protect the sample surface from gallium ion (Ga$^{+}$) beam damage. To extract the indented area in indented CeO$_{2}$/Pt/YSZ for preparing the TEM sample, the indent point was found, as shown in Figure S3a. To protect the sample surface, including the indent point from the Ga$^{+}$ ion source, 500 nm thick Pt was additionally deposited by electron beam, followed by ion beam-mediated deposition of Pt (2 $\mu$m thick) using a gas injection system (GIS), as shown in Figure S3b. The sample was milled using Ga$^{+}$ ion beam at 30 kV and at various currents. Tungsten omniprobe was used to pick up the milled sample and to attach it on a copper half TEM grid (Ted Pella, Omniprobe Lift-Out grids; Figure S3c). After sample attachment on the grid, fine milling was conducted by reducing milling current gradually (Figure S3d). Afterwards, fine milling at reduced ion energy (Ar source, 500 eV) was carried out to remove the amorphous surface layer (Fischione NanoMill 1040).

To characterize the change of oxidation state of Ce ion in the mechanically deformed CeO$_{2}$ thin film, STEM-DF imaging (JEOL 2010F) was performed. A 200 kV electron beam is focused down to a spot with probe size of 0.7 nm, and scanned across a thinned sample to electron transparency. EEL spectra were acquired in STEM mode; one spectrum for every $\approx$100 nm distance were acquired along the CeO$_{2}$ thin film. Typical acquisition times were 2 s per spectrum. In addition, the atomic structural change in the indented CeO$_{2}$ thin film was characterized by STEM-BF and STEM-HAADF imaging (Titan, FEI) operated at 300 kV with sub-Å resolution.

\subsection{Oxygen vacancy formation energy}
We modeled fluorite-structured bulk CeO$_{2}$ using a $2 \times 2 \times 2$ supercell composed of 96 atoms. To address the strongly correlated and localized 4\textit{f} shell of Ce, We employed the DFT + Hubbard \textit{U} (DFT + \textit{U}) approach. DFT + \textit{U} calculations were performed using VASP [47,48] with projected augmented wave (PAW) pseudopotentials from the VASP database and generalized gradient approximation (GGA) of Perdew-Burke-Ernzerhof (PBE) [49]. We chose \textit{U}$_{eff}$ (\textit{U-J}) = 4 eV [50] to describe the exchange-correlation effects. An energy cutoff of 600 eV and a $6 \times 6 \times 6$ Monkhorst-Pack k-point mesh were used after convergence tests. Atom positions were relaxed until all forces were less than 0.005 eV/\AA.

The perfect CeO$_{2}$ bulk phase was biaxially deformed by -10, -7.5, -5, -2.5, +2.5, +5, +7.5, and +10\% (+ means tensile strain) by changing the lattice parameters along the x- and y-axes. Along the z-axis, the corresponding deformation was applied considering Poisson’s ratio of CeO$_{2}$ [51]. To calculate the formation energy of oxygen vacancies (E$^{f}$[V$_{O}$]) in CeO$_{2}$ with respect to biaxial strain ($\epsilon$$_{x,y}$), one O atom was eliminated from the perfect CeO$_{2}$ supercells under $\epsilon$$_{x,y}$, and then relaxed as for the perfect CeO$_{2}$ supercell. The formation energy of the oxygen vacancies, E$^{f}$[V$_{O}$], with charge \textit{q} in bulk CeO$_{2}$ is defined as 

\begin{equation}
E^{f}[V_{O}^{q}] = E_{tot}[V_{O}^{q}]-E_{tot}^{perfect}+\sum{n_O\mu_O}+q(E_F + E^{perfect}_{VBM}+\Delta V_{avg}),
\end{equation}
\\
where E$_{tot}$[V$_{O}^{q}$] is the total energy of a relaxed supercell containing V$_{O}^{q}$, and E$_{tot}^{perfect}$ is the total energy for the perfect crystal using an equivalent supercell. The integer \textit{n}$_{O}$ indicates the number of O atoms that have been removed from the supercell to form the defect, and $\mu_{O}$ are the oxygen chemical potentials. The chemical potential for electrons is the Fermi energy, E$_{F}$, which is measured from the valence band edge maximum (VBM). E$_{VBM}^{perfect}$ is the VBM of the perfect supercell, which is obtained by E$_{tot}^{perfect}$ - E$_{tot}^{(perfect,+1)}$, where E$_{tot}^{(perfect,+1)}$ is the total energy of the +1 charged perfect supercell. $\Delta$V$_{avg}$ is the difference in average potentials (V$_{avg}$) far from the defect relative to the perfect supercell, i.e.,  $\Delta$V$_{avg}$ = V$_{avg}^{V_{O}^{q}}$ - V$_{avg}^{perfect}$, where V$_{avg}^{V_{O}^{q}}$ and V$_{avg}^{perfect}$ are the average potentials of the defective and perfect supercells, respectively.

\subsection{Deformation-induced oxidation state}
To dynamically and cost effectively deform CeO$_{2}$, we performed classical MD simulations using the LAMMPS package [52]. For inter-ionic potentials, a rigid ion model combining a Coulomb interaction with a short-range Buckingham term [53] was used with a cut-off radius of 10.5 \AA, by applying periodic boundary conditions along 3-D directions. Since the inhomogeneous (plastic) deformation behavior is not visibly captured if a supercell is too small, we varied the supercell size and relaxed the supercells with MD using canonical ensemble, Nos\'e-Hoover thermostat at 298 K for 100 ps (Figure S4b). Subsequently, uniaxial compressive strain ($\epsilon$$_{z}$) was imposed on the supercells at a constant rate of $5 \times 10$$^{8}$ /s at 298 K. The atomic configurations right beyond the yield point, i.e., in the plastic regime, were investigated (Figure S4c). Based on the atomic configurations, we determined the minimum size needed for conducting uniaxial compression of the CeO$_{2}$ supercell with MD simulations.

Static DFT calculations were then conducted to investigate the change in charge states of individual atoms (\textit{i}) with respect to compressive strain ($\epsilon$$_{z}$), with the same parameters used as for the bulk CeO$_{2}$ supercell explained in the previous section ‘Oxygen vacancy formation energy’. The change in atomic charge (i.e., charge difference ($\Delta\rho$) was referenced to the average charge of each snapshot under $\epsilon$$_{z}$ (i.e., $\Delta\rho$(\textit{i},$\epsilon$$_{z}$) = $\rho$(\textit{i},$\epsilon$$_{z}$) – 
$\overline{\rho}$($\epsilon_{z}$)), where total charge of \textit{i} was integrated in a constant radius (volume) from the core.

\section{Results}
\subsection{Indentation-induced current contrast change}
Figure S1a shows load ($\textit{p}$) versus displacement ($\textit{h}$) as the $\textit{p-h}$ curve of CeO$_{2}$/Pt/YSZ (YSZ = Y$_{2}$O$_{3}$ stabilized ZrO$_{2}$) recorded after indentation with a Berkovich tip. Permanent displacement of $\approx$ 100 nm in depth is observed from the indentation even after load removal. In addition, the \textit{p-h} curve of CeO$_{2}$/Pt/YSZ exhibited smooth loading-unloading without any detectable pop-in events, which is due to the presence of the ductile metal Pt and thin CeO$_{2}$ layers. The \textit{p-h} curve obtained from Pt/YSZ during indentation (Figure S1b) shows a similar aspect to that of CeO$_{2}$/Pt/YSZ, except for the reduced permanent deformation depth ($\approx$ 70 nm) in Pt/YSZ compared to CeO$_{2}$/Pt/YSZ.

To evaluate how electrical properties change due to local mechanical deformation, C-AFM scanning was performed with a tip bias of 10 V. Figures 1a--c exhibit the scanned height, deflection, and current images around an indent on CeO$_{2}$/Pt/YSZ. The C--AFM images obtained from Pt/YSZ indented under the same conditions as for CeO$_{2}$/Pt/YSZ are shown in Figures 1d--f for comparison. As can be seen in Figures 1b and 1e, the pile-up area that is known to normally be observed in ductile materials is formed by the indentation up to 1.5 -- 2 $\mu$m in lateral length both in CeO$_{2}$/Pt/YSZ and in Pt/YSZ. 

However, inhomogeneous contrast change in the current map is observed only for the CeO$_{2}$/Pt/YSZ sample (Figure 1c), while Pt/YSZ does not show any significant contrast change in the current image (Figure 1f). This clearly suggests that the current contrast change observed in CeO$_{2}$/Pt/YSZ mainly arises from the CeO$_{2}$ thin film layer. As the tip bias increases, the average contrast of the entire current map becomes darker (Figure S5), and high current flows through the indented area in a roughly triangular shape (Figure 1c). This suggests that the electrical properties, in particular electronic conductivity (refer to Figure S2), can be enhanced by local mechanical deformation. However, the degree of enhancement does not increase proportionally to the applied bias, although the average current passing through the sample is higher at a higher bias (Figure S5). In addition, the high current contrast does not exactly follow the indented pyramidal shape and pile-up configuration (Figures S6--7), which confirms that the current contrast observed in CeO$_{2}$/Pt/YSZ does not stem from a topological effect (Figure S7). A similar aspect observed in CeO$_{2}$/Pt/YSZ is also observed in the CeO$_{2}$ thin film deposited on another type of substrate (Figure S8), confirming that mechanical deformation modulates the electronic characteristics of CeO$_{2}$ thin films at room temperature.\\

\includegraphics[width=150mm]{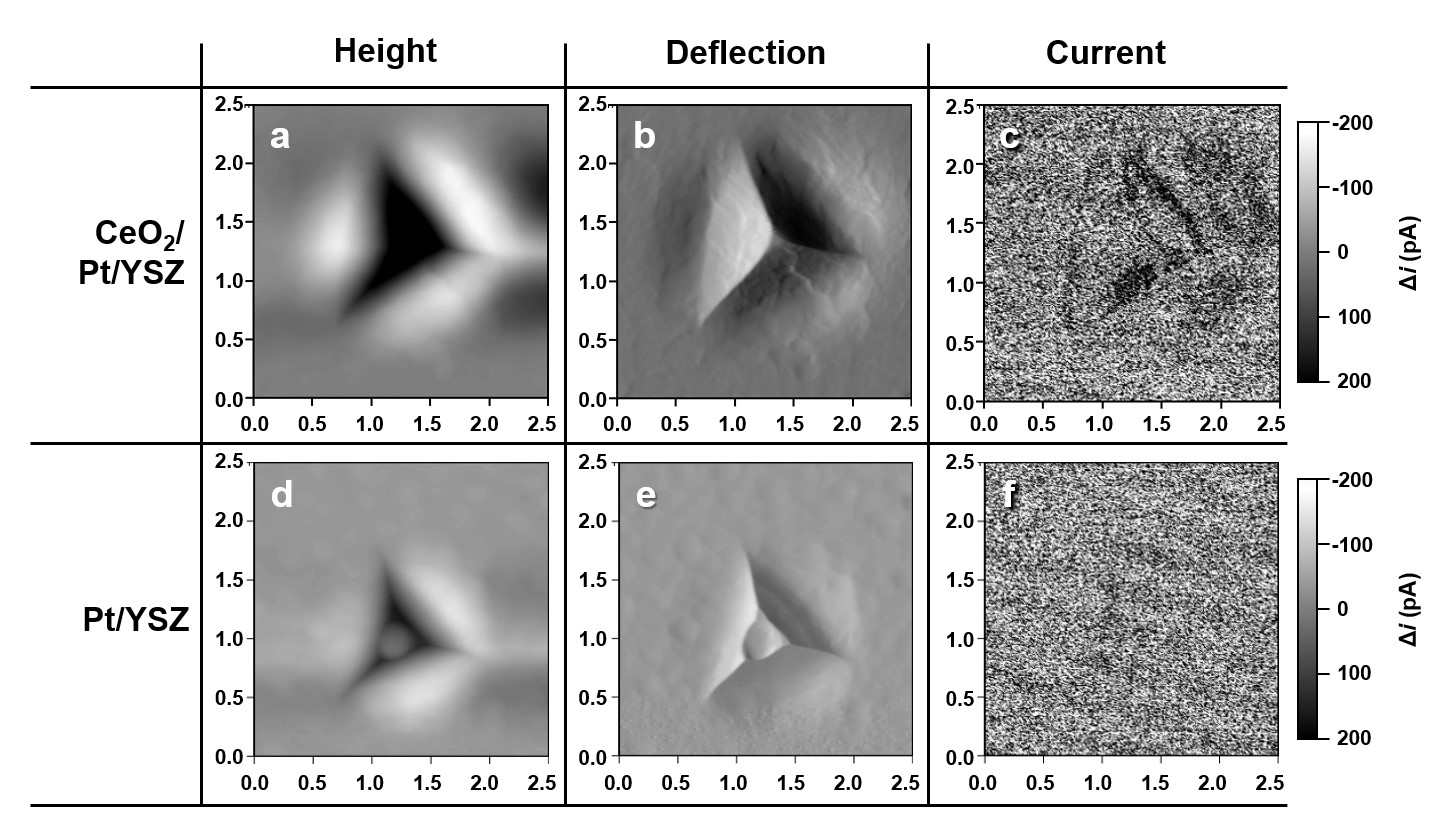}\
 
\textbf{Figure 1} \textbf{(a)} Height, \textbf{(b)} deflection and \textbf{(c)} current maps of CeO$_{2}$/Pt/YSZ indented at room temperature. \textbf{(d)} Height, \textbf{(e)} deflection and \textbf{(f)} current maps of Pt/YSZ indented at room temperature. The applied tip bias is 10 V. The AFM scanning area is $2.5 \times 2.5$ $\mu$m$^{2}$. The contrast ranges in height, deflection, and current images are 150 nm, 20 nV, and 400 pA, respectively. The dark region in the current map corresponds to a region of high current passing through CeO$_{2}$ thin film (refer to Figure S5).\\
\\
\subsection{Oxidation state change of Ce ions in the deformed area}
It was previously reported that electrical conductivity can be modified by structural changes originating from the formation of dislocation [28,29,37,38], grain boundaries [10,21], surfaces [54], and interfaces [22-24]. The common observation in the modified electrical conductivity was the presence of oxygen vacancies. In particular, it is well known that the presence of oxygen vacancies in fluorite-structured CeO$_{2}$ is accompanied by a change in the oxidation state of the cation originating from electron localization in the unoccupied 4$\textit{f}$ orbitals of two Ce$^{4+}$, resulting in the reduction of Ce ions [55]. The average oxidation state of Ce ions can be quantitatively analyzed by calculating the area of the second derivative of the extracted spectra, especially Ce-M$_{4,5}$ edges as explained in Refs. 56 and 57, using EELS by specifying a region of interest (ROI) and imaging with TEM. The relative characteristics of Ce-M$_{4,5}$ edges quantitatively reflect the electronic transition from the 3$\textit{d}$ to 4$\textit{f}$ state of Ce ions (i.e., the amount of Ce ions in valence states 3+ and 4+) [58,59].

Figures 2a and b exhibit TEM-bright field (TEM-BF) and scanning transmission electron microcopy-dark field (STEM-DF) images obtained from Berkovich tip-indented CeO$_{2}$/Pt/YSZ prepared using focused ion beam (FIB) milling (refer to Methods). Pile-up of $\approx$ 100 nm in height is generated around the indent (marked with a yellow star), the lateral width of which is similar to the pile-up width observed in C-AFM images. No evidence of cracks generated  in the indented CeO$_{2}$ layer is observed.

Figure 2c shows the Ce-M$_{4,5}$ edge EEL spectra obtained from the regions near the indent point (denoted by ‘on-indent’ in Figure 2c) and far away ($\approx$ 5 $\mu$m) from the indent point (denoted by ‘off-indent’ in Figure 2c). The two Ce-M$_{4,5}$ edges in the on- and off-indent regions display little change in the energy position; only the Ce-M5 edge in the on-indent EEL spectrum shifts by $\approx$ 0.5 eV to a lower energy loss. Figure 2d shows the Ce-M$_{5}$/M$_{4}$ ratio calculated with respect to different positions on the CeO$_{2}$ thin film. A prominent increase in the Ce-M$_{5}$/M$_{4}$ edge ratio (0.85 -- 1.02, blue circles in Figure 2d) is observed in close proximity to the pile-up region compared to that obtained from the off-indent position (Ce-M$_{5}$/M$_{4}$ ratio = 0.87, light red solid line in Figure 2d) or the value reported for bulk CeO$_{2}$ and stoichiometric grain boundaries (Ce-M$_{5}$/M$_{4}$ ratio $\approx$ 0.86 -- 0.90 [10,57,60]). This indicates that the amount of Ce$^{3+}$ has increased in the locally deformed area, suggesting that some Ce$^{4+}$ ions are reduced to Ce$^{3+}$ by mechanical deformation at room temperature. A survey of the literature reveals that the increment in the Ce-M$_{5}$/M$_{4}$ ratio in the pile-up region is analogous to those observed in nonstoichiometric grain boundaries [10,57] containing structural vacancies at the boundary (Figure 2e). Hence, the increased electronic conductivity in the indent site (Figure 1) is closely related to increased carrier concentration (i.e., n = [Ce$^{'}_{Ce}$], similar to the formation of nonstoichiometric grain boundaries or oxygen vacancies) when compared to the intact CeO$_{2}$ matrix.\\

\includegraphics[width=160mm]{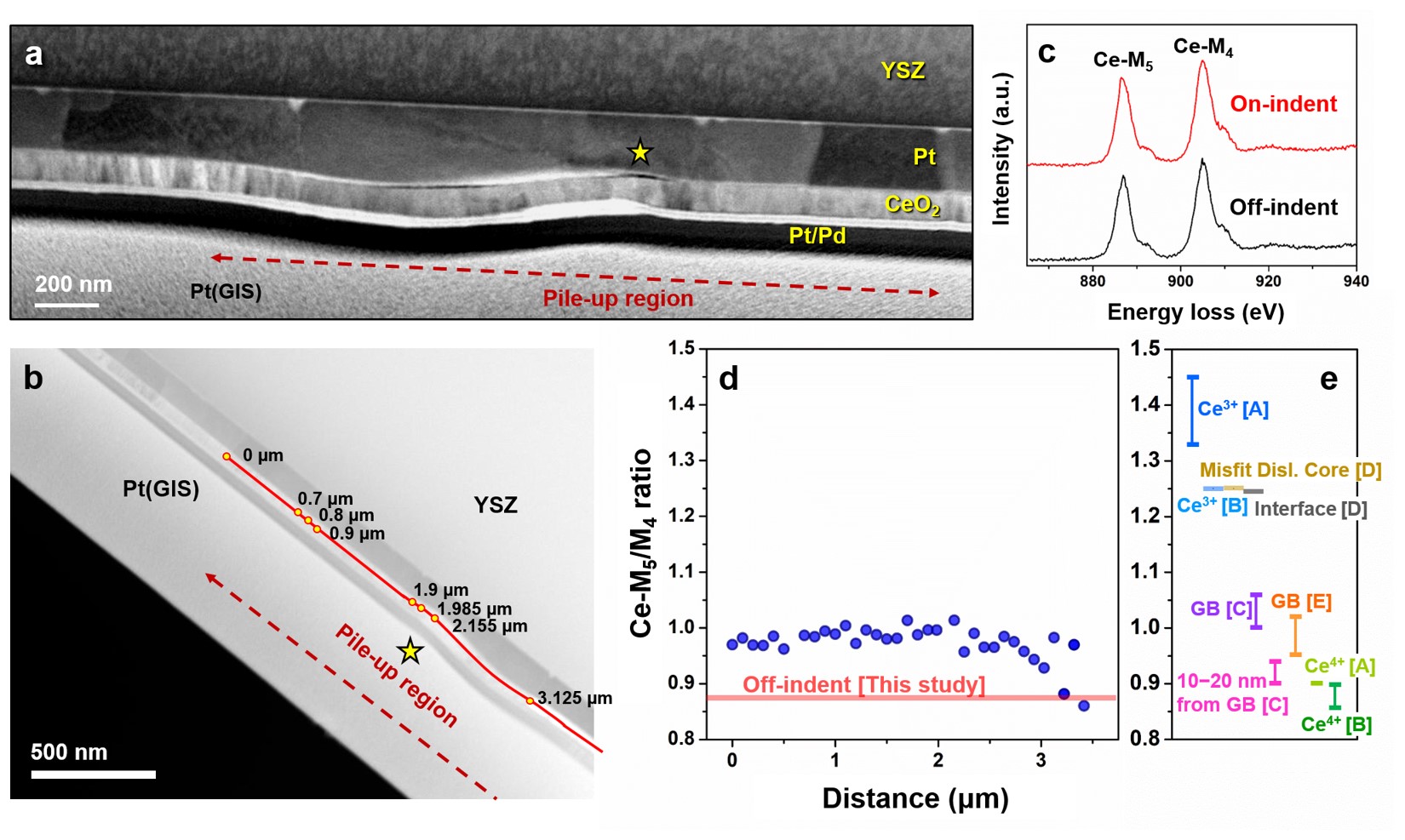}

\textbf{Figure 2} \textbf{(a)} TEM-BF and \textbf{(b)} STEM-DF images of indented CeO$_{2}$/Pt/YSZ. The region for line profiling is indicated with a red solid line on the CeO$_{2}$ layer in figure (b). The indent point and pile--up region are marked with a yellow star and a red dashed line in figures (a) and (b), respectively. \textbf{(c)} Ce M$_{4,5}$ edge EEL spectra obtained from the regions near the indent point (denoted by “on-indent”) and at a greater distance ($\approx$ 5 $\mu$m) from the indent point (denoted by “off-indent”). \textbf{(d)} Ce-M$_{5}$/Ce-M$_{4}$ intensity ratio (blue circle) obtained from STEM-EELS line profiling around the indent point (indicated with a red solid line in figure (b)). For comparison, the Ce-M$_{5}$/M$_{4}$ ratio (horizontal light red line) obtained from the region further from the indent point (denoted by “off-indent”) was superimposed in figure (d). \textbf{(e)} The Ce-M$_{5}$/M$_{4}$ ratio reported in the literature for comparison with figure (d). References [A], [B], [C], [D], and [E] in figure (e) correspond to Refs. 60, 61, 57, 56, 10, respectively. 
\\
\\
Similar to what is observed in our study, a few experimental results have shown that mechanical deformation at room temperature, such as from ball milling [62,63] and bending [64], changed the oxidation state of cations. Sergo \textit{et al.} [64] insisted that the reduction of the cations is attributed to the diffusion of oxygen (or vacancies), since the diffusion of oxygen vacancies under tensile stress at room temperature is much faster than that extrapolated from high temperature oxygen diffusivity. However, it is still questionable as to how far oxygen ions indeed diffuse in a material with very low oxygen diffusivity at room temperature (10$^{-38}$ m/s for CeO$_{2}$-Al$_{2}$O$_{3}$ [64]), even though the diffusivity is enhanced by tensile strain. Therefore, to determine the origin of the variation in oxidation state of Ce ions under mechanical deformation, in the following discussion, we investigate whether oxygen vacancies can be formed only under mechanical deformation at room temperature. 

\section{Discussion}
\subsection{Oxygen vacancy formation energy under biaxial strain}
Figure 3 shows the formation energies (E$^{f}$) of neutral, singly, and doubly charged oxygen vacancies (V$_{O}^{0}$, V$_{O}^{1+}$ and V$_{O}^{2+}$) in pristine, compressively, and tensile strained bulk CeO$_{2}$ phases  (refer to Methods). In the undistorted (pristine) CeO$_{2}$ phase, one neutral oxygen vacancy (V$_{O}^{0}$) requires 2.9 eV to be formed in the structure, which is analogous to the calculated results from previous reports [31,65,66] Under the application of in-plane biaxial strain ($\epsilon_{x,y}$), all the formation energy of V$_{O}^{0}$ is positive, regardless of the direction of strain. In addition, it decreases under tensile strain, while increasing under compressive strain, in good agreement with Refs. 31 and 37. However, the linear relationship between the formation energy of V$_{O}^{0}$ and biaxial strain is not preserved under severely compressive strain ($\epsilon_{x,y}$  $>$ 5\%), saturating at E$^{f}$= 3.5 -- 3.6 eV/vacancy, which is in good agreement with the non-monotonic variation in the formation energy of V$_{O}^{0}$ in Ref. 31. This trend observed in the formation energy of V$_{O}^{0}$ is almost the same for V$_{O}^{1+}$, although the values sometimes deviate from those observed from V$_{O}^{0}$.

However, the formation energy of V$_{O}^{2+}$ is significantly lower throughout the entire range of strain, showing that the formation energy of V$_{O}^{2+}$ is almost zero, which means that V$_{O}^{2+}$ is much more likely to be formed in CeO$_{2}$ compared to V$_{O}^{0}$ and V$_{O}^{1+}$, in accordance with the DFT calculations in Ref. 67. Therefore, it can be suggested that a predominant defect in fluorite-structured CeO$_{2}$ is V$_{O}^{2+}$, and that the doubly charged state should be considered for DFT calculations in order to appropriately predict the defect chemistry in CeO$_{2}$. In addition, the linear trend observed for the formation energies of V$_{O}^{0}$ and V$_{O}^{1+}$ is not present; under small strain (-2.5 $<$ $\epsilon_{x,y}$ $<$ 2.5 \%), the formation energy of V$_{O}^{2+}$ decreases under tensile strain, while increasing under compressive strain, in accordance with the trends observed for V$_{O}^{0}$ and V$_{O}^{1+}$. In addition, V$_{O}^{2+}$ shows the lowest formation energy with a negative value at $\epsilon_{x,y}$ $\approx$ 2.5 \% tensile strain, indicative of favorable formation of V$_{O}^{2+}$ in CeO$_{2}$ under small tensile strain even at 0 K (also at room temperature). However, for both large compressive and tensile strains, the formation energy of V$_{O}^{2+}$ increases compared to the pristine CeO$_{2}$ phase, suggesting that large in-plane strain does not help create V$_{O}^{2+}$ in the CeO$_{2}$ phase.

Considering that Poisson’s ratio ($\nu$ = $\epsilon_{x,y}$/$\epsilon_{z}$) of CeO$_{2}$ is 0.29368, when uniaxial compressive strain ($\epsilon_{z}$) of 8.53\% is applied in the height direction (along the z-axis in Figure 4b), the 2.5\% in-plane tensile strain (showing the minimum formation energy of V$_{O}^{2+}$) can also be applied to CeO$_{2}$ by uniaxial compression. This directly suggests that uniaxial compressive strain of 8.53\% applied to CeO$_{2}$ along the z-axis can create V$_{O}^{2+}$ defects in terms of the formation energy of oxygen vacancies.\\

\includegraphics[width=110mm]{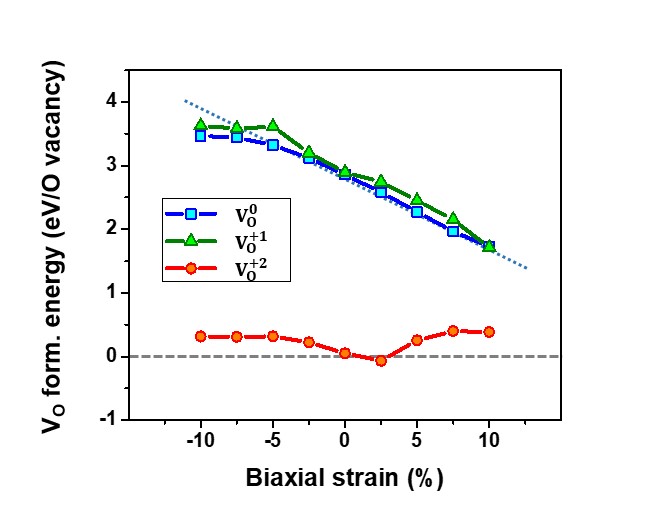}\\
\textbf{Figure 3} Formation energies of neutral (V$_{O}^{0}$), singly charged (V$_{O}^{1+}$) and doubly charged (V$_{O}^{2+}$) oxygen vacancies per vacancy in bulk CeO$_{2}$, as a function of biaxial strain ($\epsilon_{x,y}$) along [100] and [010] directions. The negative and positive strains denote compressive and tensile strains, respectively. 
\\
\\

\subsection{Local phase transformation by deformation}
To demonstrate whether oxygen vacancies are generated in the CeO$_{2}$ thin film by dynamic uniaxial compression (indentation in our experimental study) and how the transition of Ce$^{4+}$ $\rightarrow$ Ce$^{3+}$ occurs under deformation at room temperature, we performed the uniaxial compression of a perfect CeO$_{2}$ supercell without any structural defects using classical MD simulations at 298 K, as shown in Figures 4a--j. Figure 4a shows the stress-strain curve obtained during uniaxial compression of the CeO$_{2}$ supercell shown in Figure 4b. The reason for the choice of supercell size is explained in detail in Figure S6. To mimic the deformation behavior of the CeO$_{2}$ thin film under experimental indentation, the aspect ratio (thickness/width = c/a) of the CeO$_{2}$ supercell was reduced to 0.5 (Figure 4b), but using the minimum size for capturing inhomogeneous deformation during uniaxial compression, as interpreted in Figure S4. The CeO$_{2}$ supercell exhibits a linear stress-strain curve up to the yield point ($\epsilon_{z}$ $\approx$ 8.4\%), exhibiting elastic deformation with Young’s modulus of 410 GPa, which is very high compared to the experimental value (175 GPa) [51]. This is because experimental samples usually contain many defects, thus, Young’s modulus is lower compared to the theoretical value [69], suggesting a large effect from defects on the physical properties.

Beyond the yield point, the stress-strain curve is no longer linear, experiencing plastic deformation with phase transformation in a localized region, as shown in Figures 4c--e. Figure 4d shows the plastically deformed CeO$_{2}$ supercell (Figure 4c) aligned to [110]. A small region in the uniaxially compressed CeO$_{2}$ supercell is magnified and shown in Figure 4e. Uniaxial compression causes inhomogeneous deformation in the CeO$_{2}$ supercell (Figures 4b--e); the initial CeO$_{2}$ structure (left side of the CeO$_{2}$/T-CeO$_{2}$ interface; vertical blue dotted line in Figure 4e) remains in the initial CeO$_{2}$ bonding configuration (Figure 4f) in the less deformed region (four O bonds and eight Ce bonds). However, the CeO$_{2}$ structure has partially transformed into a different structure (referred to as T-CeO$_{2}$ herein) with four bonds to O, but fewer bonds to Ce (seven bonds; Figure 4e), which is consistent with those in oxygen vacancy-containing CeO$_{2}$ (Figure 4g). Considering that one of the Ce-O bond lengths is very long compared to the other three Ce-O bonds (Figure 4e), the elongated Ce-O bond with weaker bond strength is most likely to be broken during a phase transformation (Figure 4h), whereby the polyhedral structure around the central Ce ion in the T-CeO$_{2}$ phase becomes octahedral short-range ordered (SRO), with each O having three bonds and each Ce with six bonds (Figures S9a--b), differing from pristine CeO$_{2}$ (Figures S9g--h). The newly created T-CeO$_{2}$ phase via phase transformation has some structural aspects (or bonding configuration) analogous to tetragonal CeO$_{2}$ (P4$_{2}$/mnm), trigonal Ce$_{7}$O$_{12}$  (R$\overline{3}$), and cubic Ce$_{2}$O$_{3}$ (Ia$\overline{3}$) structures (Figure S9 and Table 1), thereby suggesting that the T-CeO$_{2}$ phase may exhibit electronic properties of both stoichiometric tetragonal CeO$_{2}$ and non-stoichiometric species (such as Ce$_{7}$O$_{12}$ and Ce$_{2}$O$_{3}$) that deviate from cubic CeO$_{2}$, even though T-CeO$_{2}$ remains in the same Ce:O ratio of 1:2 as pristine cubic CeO$_{2}$ without a change in stoichiometry. The change in bonding configuration occurs via the movement of oxygen ions mainly along the z- and y-axes, as marked with black arrows in Figures 4h and i, such that O ions get closer to Ce ions due to compressive strain beyond the critical point.

Indeed, a similar oxygen ion movement aspect is observed in the formation of oxygen vacancies in bulk CeO$_{2}$ (Figure S10). The formation of oxygen vacancies in CeO$_{2}$ is accompanied by miniscule migration of oxygen ions around the vacancy from their original location (Figure S10a), which in turn induces the collection of excess electron density around the oxygen vacancy (Figure S10b). Therefore, it is expected that charge redistribution near the T-CeO$_{2}$ phase is analogous to when an oxygen vacancy is generated (explained in the section ‘Charge state vs. strain’).

In short, the phase transformation is considered to occur via the following mechanism. The mechanical energy externally applied during uniaxial compression is absorbed in the CeO$_{2}$ matrix, followed by non-uniform Ce-O bond lengthening near the yield point in the elastic regime (Table S2); some Ce-O bond lengths decrease or increase a little (refer to CO2 in Table S2) compared to that of the pristine CeO$_{2}$ structure (refer to CO1 in Table S2). This occurs in order to sustain the high mechanical energy absorbed in the CeO$_{2}$ structure during uniaxial compression, by slightly distorting the bonding configuration of the CeO$_{2}$ structure. However, at the moment that the CeO$_{2}$ structure cannot sustain the severe distortion anymore (i.e., at the yield point), one of the elongated bonds is broken, while the other elongated bond becomes tighter, resulting in the creation of the T-CeO$_{2}$ phase with reduced bonding number (refer to CO4 in Table S2) by releasing absorbed mechanical energy. A part of the Ce-O bonds that do not participate in the formation of the new phase (T-CeO$_{2}$ phase) returns to the structure similar to that of pristine CeO$_{2}$ (refer to CO3 in Table S2).
The CeO$_{2}$ structure with a low aspect ratio (= 0.5) exhibits a CeO$_{2}$/T-CeO$_{2}$ phase interface parallel to the compressive strain axis (Figure 4c) during plastic deformation, unlike the CeO$_{2}$ structure with a higher aspect ratio which shows approximately 40--50 degree between the plastically deformed region (T-CeO$_{2}$) and the compressive axis (Figure S4). During uniaxial compressive deformation of the CeO$_{2}$ supercell with c/a = 1, the CeO$_{2}$/T-CeO$_{2}$ phase interface is parallel to the strain axis in the early stage of plastic deformation, while shifting 49 degree from the axis in the latter stage (Figure S11). Therefore, it should be noted that atomic rearrangement occurring during plastic deformation is largely determined by the aspect ratio of the sample, as known from other studies [70,71]. Thus, it is expected that a CeO$_{2}$/T-CeO$_{2}$ phase interface is created parallel to the loading axis in our CeO$_{2}$ thin film indentation experiment.

As can be seen in Figures 4d--h, Ce ions in the CeO$_{2}$ phase exhibit a diamond structure, albeit the bond lengths are a little shorter compared to pristine CeO$_{2}$ due to volume contraction upon compression. This diamond structure is also observed in STEM-BF and STEM-high angle annular dark field (STEM-HAADF) images near the indent on the CeO$_{2}$ thin film (Figures 4k, S12a, and S12c) and undistorted CeO$_{2}$ thin film (far removed from the indent), as shown in Figure S13. Even though the CeO$_{2}$ phase has transformed into the T-CeO$_{2}$ phase, the Ce ions in the newly created T-CeO$_{2}$ phase remain in the diamond structure, both in the uniaxially compressed CeO$_{2}$ supercell with MD simulation (Figures 4e) and in the experimentally analyzed STEM images of the indented CeO$_{2}$ thin film (Figures 4l and S12c). 

On the other hand, a difference is observed in the STEM-BF images of CeO$_{2}$ and T-CeO$_{2}$ phases in Figures 4k and l. The O ions (and open space for vacuum transmission) have diamond symmetry along the [110] zone axis of the CeO$_{2}$ phase (Figures 4k, S12a, and S13) as expected from Figures 4d--f, whereas bright contrast in a zigzag pattern (not straight diamond symmetry) along the diagonal direction is observed for the T-CeO$_{2}$ phase (Figure 4l and (2-5) in Figure S12). This difference is thought to arise from the broken symmetry of O ions (including the contrast coming from vacuum transmission) by the movement of O ions in the T-CeO$_{2}$ phase (as illustrated in (2-3) to (2-5) in Figure S12). This zigzag pattern observed along the diagonal direction has never been observed from the cubic structure of CeO$_{2}$, and we could not observe this peculiar contrast in the undistorted region of the CeO$_{2}$ thin film (Figure S13). Thus, it appears that the T-CeO$_{2}$ phase formed via mechanical deformation differs from the cubic CeO$_{2}$ structure (Figures 4l and S12c).\\

\includegraphics[width=160mm]{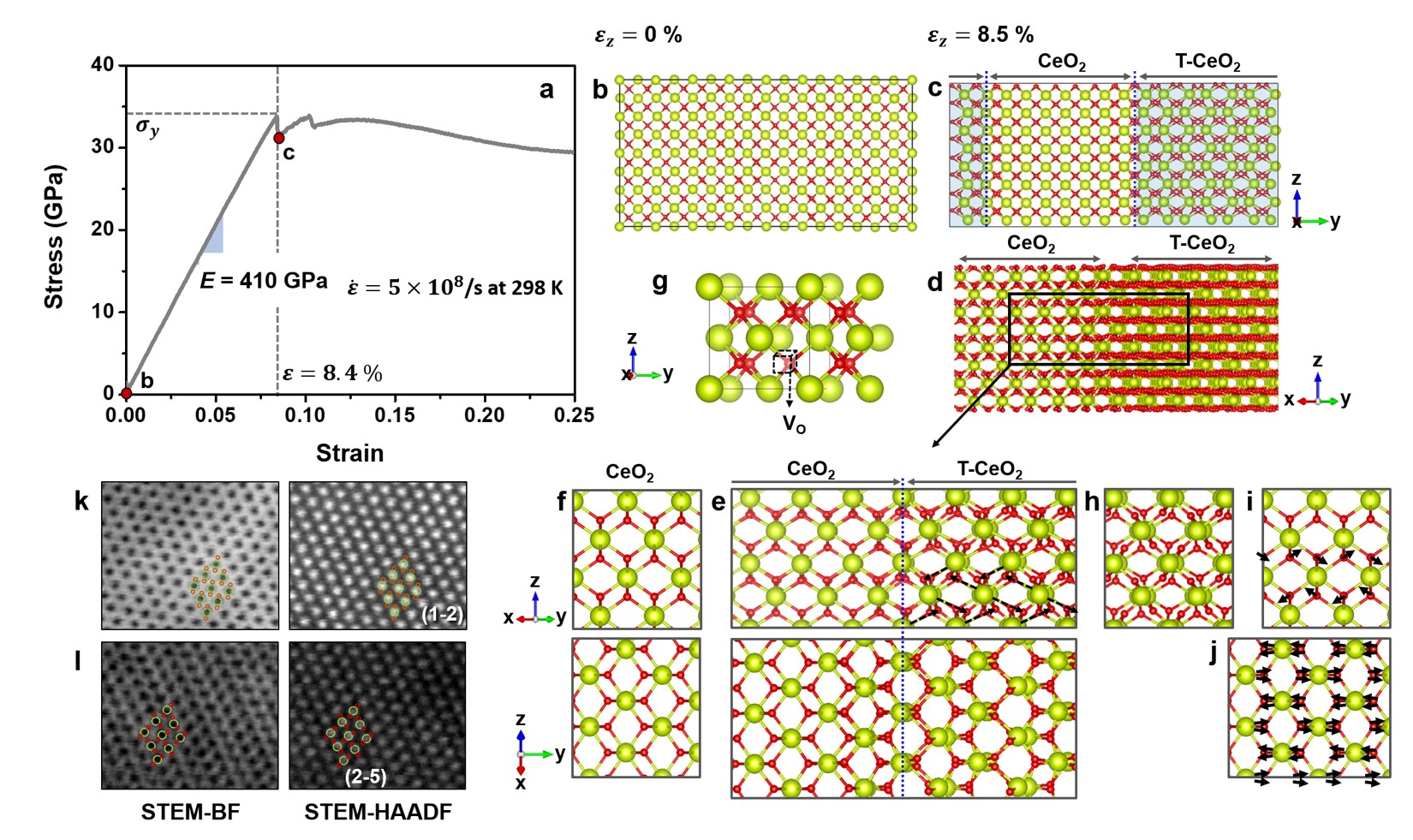}\\
\textbf{Figure 4} \textbf{(a)} Stress--strain curve obtained during uniaxial compression along height direction (z--axis), shown in figure (b). \textbf{(b)} Initial structure of CeO$_{2}$ supercell with aspect ratio (c/a) = 0.5, before uniaxial compression. Yellow--green and red circles denote Ce and O atoms, respectively. \textbf{(c)} Atomic configuration of the CeO$_{2}$ supercell uniaxially compressed right beyond the elastic limit, under $\epsilon_{z}$ = 8.5\%. The phase transformed region (CeO$_{2}$ $\rightarrow$ T-CeO$_{2}$) is highlighted by the blue shade. The CeO$_{2}$/T-CeO$_{2}$ interface is denoted with blue dotted lines. \textbf{(d)} Atomic configuration of the plastically deformed CeO$_{2}$ supercell ($\epsilon_{z}$ = 8.5\%, as shown in figure (c)) aligned to [110]. \textbf{(e)} The magnified view of figure (d) showing both CeO$_{2}$ and T-CeO$_{2}$ phases simultaneously. The upper figures are aligned to [110], while bottom figures are aligned to [101]. \textbf{(f)} Atomic configuration of the pristine CeO$_{2}$ aligned to [110] (upper) and [101] (bottom), shown to compare to figure (e). \textbf{(g)} Atomic structure of a CeO$_{2}$ unit cell containing an oxygen vacancy (V$_{O}$). \textbf{(h)} Atomic configuration of the T-CeO$_{2}$ phase when the highly elongated Ce-O bond in figure (e) is considered to be broken. \textbf{(i) -- (j)} O ions movement indicated with black arrows in T-CeO$_{2}$ phase aligned to [110] (figure (i)) and to [101] (figure (j)), anticipated by comparing the T-CeO$_{2}$ region in figure (e) with the pristine CeO$_{2}$ structure in figure (f). \textbf{(k) -- (l)} STEM--BF and STEM--HAADF images of the CeO$_{2}$ phase (figure (k)) and T-CeO$_{2}$ phase (figure (l)) taken from the indented CeO$_{2}$ thin film shown in Figures 2a--b, along the [110] zone axis. The detailed interpretation of the STEM images and the denotation of (1-2) and (2-5) in figures (k) and (l) are in Figure S12. In STEM--BF images, the darker contrast is the Ce ion, while the brightest contrast is vacuum, and vice versa in STEM--HAADF images.   
\\
\\

\subsection{Charge state vs. strain}
To determine the origin of the higher concentration of Ce$^{3+}$ and higher current in the indented CeO$_{2}$ thin film (Figures 1 and 2), we investigated the charge redistribution in CeO$_{2}$ during mechanical deformation. Figure 5 shows the charge difference relative to the average atomic charge of each configuration (refer to Methods) with respect to the applied uniaxial compressive strain using the MD-compressed CeO$_{2}$ supercell (Figure 4). Within the elastic regime (up to $\epsilon_{z}$ = 8.4\%, Figures 5a and b), atomic charge does not change considerably in the entire supercell, whereas it suddenly exhibits a large change just beyond the elastic limit ($\epsilon_{z}$ = 8.5\%) in Figure 5c. The charge difference in the plastically deformed CeO$_{2}$ is inhomogeneous, such that in the local area in which more severe plastic deformation occurs (T-CeO$_{2}$ phase), Ce ions tend to attract more electrons. The increased charge in the T-CeO$_{2}$ phase appears to come from the less deformed CeO$_{2}$ region, considering that the CeO$_{2}$ phase loses electrons (blue color) after plastic deformation (Figure 5c). This indicates that the plastic deformation-induced structural change from CeO$_{2}$ to T-CeO$_{2}$ causes the redistribution of electrons at and near the interface such that electrons migrate from CeO$_{2}$ to T-CeO$_{2}$. This redistribution is expected to be driven by the structural difference between the T-CeO$_{2}$ structure and the pristine CeO$_{2}$. Note that this does not involve oxygen ion movement between the two phases, and the amount of oxygen in each phase is preserved.

The oxidation state of Ce ions in the T-CeO$_{2}$ phase is best represented as Ce$_{plastic\:regime}^{\alpha+}$ (where 3 $<$ $\alpha$ $<$ 4), where the degree of electron transfer from CeO$_{2}$ to T-CeO$_{2}$ phases is approximately 0.20 -- 0.25 \textit{e} (changing from approximately -0.13 \textit{e} in the CeO$_{2}$ phase to $\approx$ 0.1 \textit{e} in the T-CeO$_{2}$ phase) in the supercell of the CeO$_{2}$/T-CeO$_{2}$ heterostructure (Figure 4c). The change is very close to the change analyzed by TEM-EELS in Figures 2d--e, considering that the reduction from Ce$^{4+}$ to Ce$^{3+}$ is a 1\textit{e} change, Ce-M$_{5}$/M$_{4}$ ratio changes from 0.87 to 1.4 ($\Delta$Ce-M$_{5}$/M$_{4}$ ratio = 0.53), and in changing from Ce$^{4+}$ to Ce$_{plastic\:regime}^{\alpha+}$, the Ce-M$_{5}$/M$_{4}$ ratio varies from 0.87 to 0.98 ($\Delta$Ce-M$_{5}$/M$_{4}$ ratio = 0.11). The same trend in the charge difference distribution in the mechanically deformed CeO$_{2}$ supercell is also observed in a CeO$_{2}$ supercell with a higher aspect ratio (Figure S14). This aspect of electron transfer occurring in mechanically deformed CeO$_{2}$ supports the results from high temperature X-ray photoelectron spectroscopy (XPS) showing that the ratio of Ce$^{3+}$/Ce$^{4+}$ in CeO$_{2}$ improved upon in-plane biaxial tensile strain [31]. Therefore, the enhanced current observed in the C-AFM analysis of the indented CeO$_{2}$ thin film (Figure 1) is thought to originate from formation of the highly electronically conductive T-CeO$_{2}$ region, which redistributes charges between CeO$_{2}$ and T-CeO$_{2}$ phases (considered as the reduction of Ce$^{4+}$). The redistribution causes a prominent increase in the electronic conductivity at relatively low temperature due to faster transport of polarons [72], resulting in enhancement of electronic conductivity of the CeO$_{2}$ thin film.\\
\includegraphics[width=160mm]{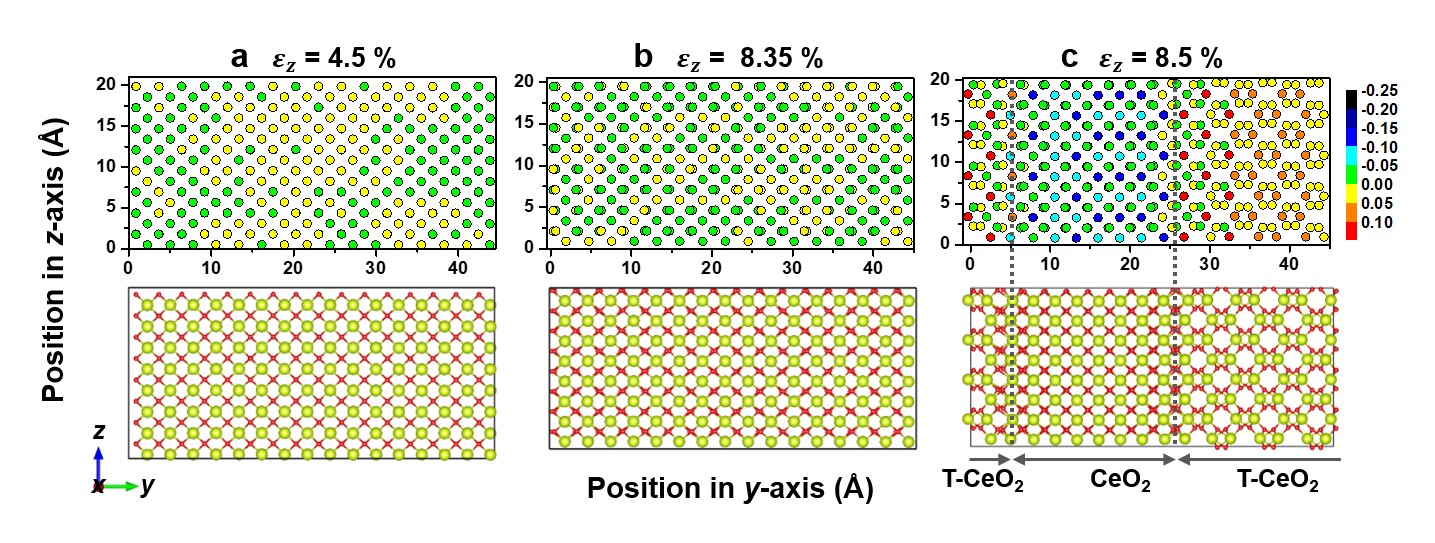}\\
\textbf{Figure 5} The distribution of charge difference (upper), and atomic configurations (bottom), in CeO$_{2}$ supercells with respect to the applied compressive strains of \textbf{(a)} 6.5\%, \textbf{(b)} 8.35\% and \textbf{(c)} 8.5\%, along the z--axis. In the upper figures, one circle denotes the position of an atom. The color map (right side of figure (c)) is identical for figures (a) -- (c), and the unit is e/atom. For the definition of the charge difference, refer to Methods. In the bottom figures, yellow-green and red circles denote Ce and O atoms, respectively. The vertical gray dotted lines are the CeO$_{2}$/T-CeO$_{2}$ interfaces.
\\
\\

\section{Conclusions}
We report that mechanical indentation at room temperature improves local electronic conductivity in CeO$_{2}$ thin films (Figure 1). The improved conductivity is closely related to the increased concentration of Ce$^{3+}$ in the indented CeO$_{2}$ matrix (Figure 2). The incremental change in concentration of Ce$^{3+}$ is similar to that observed in non-stoichiometric grain boundaries [10,57], which is known to be accompanied by accommodation of oxygen vacancies in the boundaries. 

According to our DFT calculations (Figure 3), doubly charged oxygen vacancies (V$_{O}^{2+}$) are thermodynamically favorable to be formed in CeO$_{2}$ under small biaxial tensile strain of 2.5\% along in-plane (or uniaxial compressive strain of 8.53\% along out-of-plane) directions at room temperature (or 0 K) in terms of the formation energy of oxygen vacancies. On the contrary, singly charged and neutral oxygen vacancies are not likely to be created even under a large amount of tensile strain. This implies that doubly charged oxygen vacancies can be generated during dynamic mechanical deformation, such as uniaxial compression or indentation.
Uniaxial compression with classical MD simulations and static DFT calculations of the snapshots obtained from the MD compression (Figures 4 and 5) show that plastic deformation (beyond the elastic limit) partially induces the phase transformation from CeO$_{2}$ to T-CeO$_{2}$. The phase transformation mainly takes place via movement of the O ions closer to Ce ions during uniaxial compression, which makes the bonding configuration of the T-CeO$_{2}$ phase similar to the oxygen vacancy-containing CeO$_{2}$ structure. During phase transformation, electrons redistribute such that some electrons migrate from the pristine CeO$_{2}$ to the adjacent severely deformed T-CeO$_{2}$ phase region; this results in Ce ions in the T-CeO$_{2}$ phase mainly changing oxidation state from Ce$^{4+}$ to Ce$^{\alpha+}$ (where 3 $<$ $\alpha$ $<$ 4) (Figure 5). The excess electron accumulation in Ce ions in the T-CeO$_{2}$ phase is analogous to the increment observed in the Ce-M$_{5}$/M$_{4}$ ratios of the indented CeO$_{2}$ thin film compared to that of the undistorted CeO$_{2}$ thin film, as identified by TEM-EELS (Figure 2). 
Our findings based on the use of strain engineering at room temperature are different from the previous explanation that is widely discussed to clarify the origin of the varied electrical properties by strain engineering at elevated temperature. Our study also suggests a simple method for tuning the electronic conductivity via structural change by mechanical deformation; phase transformation induced by mechanical deformation at room temperature can provide an effect similar to the introduction of oxygen vacancies, enhancing electronic conductivity in a localized region. We expect that this mechanical method can be easily employed to make heterogeneous structures that have different phases and electronic characteristics while being formed from the same constituents.

\section{ACKNOWLEDGEMENTS}
We thank Young Woo Jeong and Min Kyung Cho for help with TEM sampling with FIB-SEM and STEM imaging with TEM. This work was supported by the KIST project (2E29340).

\section{Reference}
[1] M.C. Göbel, G. Gregori and J. Maier. Numerical calculations of space charge layer effects in nanocrystalline ceria. Part II: detailed analysis of the space charge layer properties. \textit{Phys. Chem. Chem. Phys.:16.10175--10186}. 2014.\

[2] S. Li, Y. Zhu, D. Min and G. Chen.Space charge modulated electrical breakdown. \textit{Sci. Rep.:6.32588}. 2016.\

[3] K. Gelderman, L. Lee and S.W. Donne. Flat-band potential of a semiconductor: using the Mott–Schottky equation. \textit{J. Chem. Educ.:84.685--688}. 2007.\

[4] W.C. Tian, Y. H. Ho, C.H. Chen and C.Y. Kuo. Sensing performance of precisely ordered TiO$_{2}$ nanowire gas sensors fabricated by electron-beam lithography. \textit{Sensors:13.865--874} 2013.\

[5] Y.F. Sun, S.B. Liu, F.L. Meng, J.Y. Liu, Z. Jin, L.T. Kong and J.H. Liu. Metal oxide nanostructures and their gas sensing properties: a review. \textit{Sensors:12.2610--2631} 2012.\

[6] C. Wang, L. Yin, L. Zhang, D. Xiang and R. Gao. Metal oxide gas sensors: sensitivity and influencing factors. \textit{Sensors:10.2088--2106} 2010.\

[7] X. Xu, K. Takada, K. Watanabe, I. Sakaguchi, K. Akatsuka, B. T. Hang, T. Ohnishi and T. Sasaki. Self-organized core–shell structure for high-power electrode in solid-state lithium batteries.
\textit{Chem. Mater.:23.3798--3804} 2011.\

[8] J. Haruyama, K. Sodeyama, L. Han, K. Takada and Y. Tateyama. Space--charge layer effect at interface between oxide cathode and sulfide electrolyte in all--solid--state lithium--ion battery. \textit{Chem. Mater.:26.4248--4255} 2014.\

[9] B. Wu, S. Wang, W.J. Evans, D.Z. Deng, J. Yang and J. Xiao. Interfacial behaviours between lithium ion conductors and electrode materials in various battery systems. \textit{J. Mater. Chem. A:4.15266--15280} 2016.\

[10] B. Feng, I. Sugiyama, H. Hojo, H. Ohta, N. Shibata and Y. Ikuhara. Atomic structures and oxygen dynamics of CeO$_{2}$ grain boundaries, \textit{Sci. Rep.:6.20288} 2016.\

[11] Y. Lin, S. Fang, D. Su, K.S. Brinkman and F. Chen.
 Enhancing grain boundary ionic conductivity in mixed ionic–electronic conductors. \textit{Nat. Commun.:6.6824} 2015.\

[12] Y.J. Choi, I.S. Hwang, J.G. Park, K.J. Choi, J.H. Park and J.H. Lee. Novel fabrication of an SnO$_{2}$ nanowire gas sensor with high sensitivity. \textit{Nanotechnology:19.095508} 2008.\ 

[13] ]W. Zeng, T. Liu, Z. Wang, S. Tsukimoto, M. Saito and Y. Ikuhara. Selective detection of formaldehyde gas using a Cd-doped TiO$_{2}$-SnO$_{2}$ sensor. \textit{Sensors:9.9029--9038} 2009.\

[14] M.T. Greiner, M.G. Helander, W.-M. Tang, Z.-B. Wang, J. Qiu and Z.-H. Lu. Universal energy-level alignment of molecules on metal oxides. \textit{Nat. Mater.:11.76--81} 2012.\ 

[15] R. Memming. Photoinduced charge transfer processes at semiconductor electrodes and particles. \textit{Electron Transfer I:169.105--181} 1994.\

[16] A.J. Nozik. Photoelectrochemistry: Applications to solar energy conversion. \textit{Ann. Rev. Phys. Chem.:29.189--222} 1978.\ 

[17] K.W. Park and A.M. Kolpak. Optimal methodology for explicit solvation prediction of band edges of transition metal oxide photocatalysts. \textit{Commun. Chem.:2.79} 2019.\

[18] R. Chen, F. Fan, T. Dittrich and C. Li. Imaging photogenerated charge carriers on surfaces and interfaces of photocatalysts with surface photovoltage microscopy. \textit{Chem. Soc. Rev.:47.8238--8262} 2018.\

[19] Z. Zhang and J.T. Yates, Jr. Band Bending in Semiconductors: Chemical and Physical Consequences at Surfaces and Interfaces. \textit{Chem. Rev.:112.5520--5551} 2012.\

[20] K.W. Park and A.M. Kolpak.
Mechanism for spontaneous oxygen and hydrogen evolution reactions on CoO nanoparticles. \textit{J. Mater. Chem. A:7.6708--6719} 2019.\

[21] S.K. Tiku and F.A. Kröger. Effects of Space Charge, Grain-Boundary Segregation, and Mobility Differences Between Grain Boundary and Bulk on the Conductivity of Polycrystalline A1$_{2}$O$_{3}$. \textit{J. Am. Ceram. Soc.:63.183--189} 1980.\

[22] M. Basletic, J.-L. Maurice, C. Carrétéro, G. Herranz, O. Copie, M. Bibes, É. Jacquet, K. Bouzehouane, S. Fusil and A. Barthélémy. Mapping the spatial distribution of charge carriers in LaAlO$_{3}$/SrTiO$_{3}$ heterostructures. \textit{Nat. Mater.:7.621--625} 2008.\

[23] N. J. Dudney. Effect of Interfacial Space-Charge Polarization on the Ionic Conductivity of Composite Electrolytes. \textit{J. Am. Ceram. Soc.:68.538--545} 1985.\

[24] N. Sata, K. Eberman, K. Eberl and J. Maier.  Mesoscopic fast ion conduction in nanometre-scale planar heterostructures. \textit{Nature:408.946--949} 2000.\

[25] Z. Xia and S. Guo. Strain engineering of metal-based nanomaterials for energy electrocatalysis. \textit{Chem. Soc. Rev.:48.3265--3278} 2009.\ 

[26] ]T.N. Pingel, M. Jørgensen, A.B. Yankovich, H. Grönbeck and E. Olsson. Influence of atomic site-specific strain on catalytic activity of supported nanoparticles.
\textit{Nat. Commun.:9.2722} 2018.\

[27] E. Laredo, N. Suarez, A. Bello, M. Puma and D. Figueroa. Dislocation polarization and space-charge relaxation in solid solutions Ba$_{1-x}$La$_{x}$F$_{2+x}$. \textit{Phys. Rev. B:32.8325--8331} 1985.\ 

[28] K.K. Adepalli, M. Kelsch, R. Merkle and J. Maier.
 Influence of line defects on the electrical properties of single crystal TiO$_{2}$. \textit{Adv. Funct. Mater.:23.1798--1806} 2013.\ 

[29] K.K. Adepalli, J. Yang, J. Maier, H.L. Tuller and B. Yildiz. Tunable Oxygen Diffusion and Electronic Conduction in SrTiO$_{3}$ by Dislocation-Induced Space Charge Fields. \textit{Adv. Funct. Mater.:27.1700243} 2017.\

[30] M. Ahn, J. Cho and W. Lee. One-step fabrication of composite nanofibers for solid oxide fuel cell electrodes.
\textit{J. Power Sources:434.226749} 2019.

[31] C.B. Gopal, M. García-Melchor, S.C. Lee, Y. Shi, A. Shavorskiy, M. Monti, Z. Guan, R. Sinclair, H. Bluhm, A. Vojvodic and W.C. Chueh. Equilibrium oxygen storage capacity of ultrathin CeO$_{2-\delta}$ depends non--monotonically on large biaxial strain. \textit{Nat. Commun.:8.15360} 2017.\

[32] C. Zhu, X. Niu, Y. Fu, N. Li, C. Hu, Y. Chen, X. He, G. Na, P. Liu, H. Zai, Y. Ge, Y. Lu, X. Ke, Y. Bai, S. Yang, P. Chen, Y. Li, M. Sui, L. Zhang, H. Zhou and Q. Chen. Strain engineering in perovskite solar cells and its impacts on carrier dynamics. \textit{Nat. Commun.:10.815} 2019.\

[33] B. You, M.T. Tang, C. Tsai, F. Abild-Pedersen, X. Zheng and H. Li. Enhancing Electrocatalytic Water Splitting by Strain Engineering. \textit{Adv. Mater.:31.1807001} 2019.\

[34] A. Castellanos-Gomez, R. Roldan, E. Cappelluti, M. Buscema, F. Guinea, H.S.J. van der Zant and G.A. Steele. Local Strain Engineering in Atomically Thin MoS$_{2}$. \textit{Nano Lett.:13.5361--5366} 2013.\

[35] J.F. Wu and X. Guo. Origin of the low grain boundary conductivity in lithium ion conducting perovskites: Li$_{3x}$La$_{0.67x}$TiO$_{3}$. \textit{Phys. Chem. Chem. Phys.:19.5880--5887} 2017.\

[36] J. Hao, J. Zheng, F. Ling, Y. Chen, H. Jing, T. Zhou, L. Fang and M. Zhou. Strain-engineered two-dimensional MoS$_{2}$ as anode material for performance enhancement of Li/Na-ion batteries. \textit{Sci. Rep.:8.2079} 2018.\

[37] L. Sun, D. Marrocchelli and B. Yildiz. Edge dislocation slows down oxide ion diffusion in doped CeO$_{2}$ by segregation of charged defects. \textit{Nat. Commun.:6.6294} 2015.\ 

[38] S.P. Waldow and R.A. De Souza. Computational Study of Oxygen Diffusion along a [100] Dislocations in the Perovskite Oxide SrTiO$_{3}$. \textit{ACS Appl. Mater. Inter.:8.12246--12256} 2016.\

[39] T.X.T. Sayle, S.C. Parker and D.C. Sayle. Ionic conductivity in nano-scale CeO$_{2}$/YSZ heterolayers.
\textit{J. Mater. Chem.:16.1067--1081} 2016.\

[40] J. Wang, F. Hu, Y. Zhao, Y. Liu, R. Wu, J. Sun and B. Shen. Effect of epitaxial strain on small-polaron hopping conduction in Pr$_{0.7}$(Ca$_{0.6}$Sr$_{0.4}$)$_{0.3}$MnO$_{3}$ thin films. \textit{Appl. Phys. Lett.:106.102406} 2015.\

[41] M. Grünbacher, L. Schlicker, M.F. Bekheet, A. Gurlo, B. Klötzer and S. Penner. H$_{2}$ reduction of Gd- and Sm-doped ceria compared to pure CeO$_{2}$ at high temperatures: effect on structure, oxygen nonstoichiometry, hydrogen solubility and hydroxyl chemistry. \textit{Phys. Chem. Chem. Phys.:20.22099} 2018.\

[42] T. Montini, M. Melchionna, M. Monai and P. Fornasiero. Fundamentals and Catalytic Applications of CeO$_{2}$--Based Materials. \textit{Chem. Rev.:116.5987--6041} 2016.\ 

[43] S.B. Khan and K. Akhtar. Doped Ceria for Solid Oxide Fuel Cells. \textit{Cerium Oxide : Applications and Attributes} pages 44--48. IntechOpen, 2019.\

[44] B. Choudhury, P. Chetri and A. Choudhury. Oxygen defects and formation of Ce$^{3+}$ affecting the photocatalytic performance of CeO$_{2}$ nanoparticles.
\textit{RSC Adv.:4.4663--4671} 2014.\

[45] Y.P. Lan and H.Y. Sohn. Effect of oxygen vacancies and phases on catalytic properties of hydrogen--treated nanoceria particles. \textit{Mater. Res. Express:5.035501} 2018.\

[46] P. Dutta, S. Pal, M.S. Seehra, Y. Shi, E.M. Eyring and R.D. Ernst. Concentration of Ce$^{3+}$ and Oxygen Vacancies in Cerium Oxide Nanoparticles. \textit{Chem. Mater.:18.5144--5146} 2006.\

[47] G. Kresse and J. Hafner. Ab initio molecular dynamics for liquid metals. \textit{J. Phys. Rev. B:47.558--561} 1993.\

[48] G. Kresse and J. Hafner. Ab initio molecular-dynamics simulation of the liquid--metal--amorphous--semiconductor transition in germanium. \textit{Phys. Rev. B:49.14251--14269} 1994.\

[49] J.P. Perdew, K. Burke and M. Ernzerhof. Generalized Gradient Approximation Made Simple. \textit{Phys. Rev. Lett.:77.3865--3868} 1996.\

[50] C. Loschen, J. Carrasco, K.M. Neyman and F. Illas. First-principles LDA+$\textit{U}$ and GGA+$\textit{U}$ study of cerium oxides: Dependence on the effective $\textit{U}$ parameter. \textit{Phys. Rev. B:75.035115} 2007.\

[51] K. Sato, H. Yugami and T. Hashida. Effect of rare-earth oxides on fracture properties of ceria ceramics. \textit{J. Mater. Sci.:39.5765--5770} 2004.\

[52] S. Plimpton. Fast Parallel Algorithms for Short-Range Molecular Dynamics. \textit{J. Comp. Phys.:117.1--19} 1995.\

[53] L. Minervini, M.O. Zacate and R.W. Grimes. Defect cluster formation in M$_{2}$O$_{3}$-doped CeO$_{2}$. \textit{Solid State Ionics:116.339--349} 1999.\

[54] G. Heiland. Surface conductivity of semiconductors and its variation by adsorption, transverse electric fields and irradiation. \textit{Discuss. Faraday Soc.:28.168--182} 1959.\

[55] M. Alaydrus, M. Sakaue and H. Kasai. A DFT+$\textit{U}$ study on the contribution of 4$\textit{f}$ electrons to oxygen vacancy formation and migration in Ln-doped CeO$_{2}$. \textit{Phys. Chem. Chem. Phys.:18.12938--12946} 2016.\

[56] K. Song, H. Schmid, V. Srot, E. Gilardi, G. Gregori, K. Du, J. Maier and P.A. van Aken. Cerium reduction at the interface between ceria and yttria--stabilised zirconia and implications for interfacial oxygen non--stoichiometry. \textit{APL Mater.:2.032104} 2014.\

[57] H. Hojo, T. Mizoguchi, H. Ohta, S.D. Findlay, N. Shibata, T. Yamamoto and Y. Ikuhara. Atomic Structure of a CeO$_{2}$ Grain Boundary: The Role of Oxygen Vacancies. \textit{Nano Lett.:10.4668--4672} 2010.\

[58] H. Dexpert, R.C. Karnatak, J.M. Esteva, J.P. Connerade, M. Gasgnier, P.E. Caro and L. Albert. X-ray absorption studies of CeO$_{2}$, PrO$_{2}$, and TbO$_{2}$. II. Rare-earth valence state by LIII absorption edges.
\textit{Phys. Rev. B:36.1750--1753} 1987.\

[59] C.M. Sims, R.A. Maier, A.C. Johnston-Peck, J.M. Gorham, V.A. Hackley and B.C. Nelso. Approaches for the quantitative analysis of oxidation state in cerium oxide nanomaterials.\textit{Nanotechnology:30.\\085703} 2019.\

[60] X. Hao, A. Yoko, C. Chen, K. Inoue, M. Saito, G. Seong, S. Takami, T. Adschiri and Y. Ikuhara. Atomic-Scale Valence State Distribution inside Ultrafine CeO$_{2}$ Nanocubes and Its Size Dependence. \textit{Small:14.1802915} 2018.\

[61] J.A. Fortner and E.C. Buck.
The chemistry of the light rare‐earth elements as determined by electron energy loss spectroscopy. \textit{Appl. Phys. Lett.:68.3817--3819} 1996.\  

[62] D.S. Su, V. Roddatis, M. Willinger, G. Weinberg, E. Kitzelmann, R. Schlögl and H. Knözinger. Tribochemical modification of the microstructure of V$_{2}$O$_{5}$.
\textit{Catal. Lett.:74.169--175} 2001.\ 

[63] A.A. Shubin, O.B. Lapina, E. Bosch, J. Spengler and H. Knözinger. Effect of Milling of V$_{2}$O$_{5}$ on the Local Environment of Vanadium as Studied by Solid-State 51V NMR and Complementary Methods. \textit{J. Phys. Chem. B:103.3138--3144} 1999.\

[64] V. Sergo, C. Schmid and S. Meriani. Mechanically induced zone darkening of alumina/ceria--stabilized zirconia composites. \textit{J. Am. Ceram. Soc.:77.2971--2976} 1994.\

[65] D.A. Andersson, S.I. Simak, N.V. Skorodumova, I.A. Abrikosov and B. Johansson. Redox properties of CeO$_{2}$--MO$_{2}$ (M=Ti, Zr, Hf, or Th) solid solutions from first principles calculations. \textit{Appl. Phys. Lett.:90.031909} 2007.\ 

[66] Z. Yang, G. Luo, Z. Lu and K. Hermansson. Oxygen vacancy formation energy in Pd--doped ceria: A DFT+$\textit{U}$ study. \textit{J. Chem. Phys.:127.074704} 2007.\

[67] T. Zacherle, A. Schriever, R.A. De Souza and M. Martin. Ab initio analysis of the defect structure of ceria. \textit{Phys. Rev. B:87.134104} 2013.\

[68] J.C. Goldsby. Basic Elastic Properties Predictions of Cubic Cerium Oxide Using First-Principles Methods. \textit{J. Ceramics:2013.323018} 2013.\

[69] H.S. Kim and M.B. Bush. The effects of grain size and porosity on the elasticmodulus of nanocrystalline materials. \textit{Nanostruct. Mater.:11.361--367} 1999.\

[70] Z.F. Zhang, H. Zhang, X.F. Pan, J. Das and J. Eckert.
Effect of aspect ratio on the compressive deformation and fracture behaviour of Zr--based bulk metallic glass.
\textit{Phil. Mag. Lett.:85.513--521} 2005.\

[71] J. Jin, J. Cao, S. Zhou, P. Yang and Z. Guo. Quasicontinuum simulations of geometric effect on onset plasticity of nano--scale patterned lines. \textit{Modelling Simul. Mater. Sci. Eng.:25.065012} 2017.\

[72] Y.-M. Chiang, E.B. Lavik and D.A. Blom.
Defect thermodynamics and electrical properties of nanocrystalline oxides: pure and doped CeO$_{2}$. \textit{Nanostruct Mater.:9.633--642} 1997.\\
\\
\\
\\
\\
\newpage
\LARGE{\textbf{Supporting information}}
\\
\\
\large{\textbf{Deformation--induced charge redistribution in CeO$_{2}$ thin film at room temperature}}\\
\\
Kyoung--Won Park$^{a,*}$, Chang Sub Kim$^{b}$
\\
\\
$^{a}$ Center for Biomaterials, Korea Institute of Science and Technology, Seoul, 02792, Republic of Korea.\

$^{b}$ Department of Materials Science and Engineering, Massachusetts Institute of Technology, Cambridge, Massachusetts 02139, USA.\\
\\

\large{\textbf{Indentation experiments}}\\
\\
Indentation experiments were performed at room temperature using a Tribo-indenter (Hysitron Inc., Minneapolis, MN) with a Berkovich tip (150 nm average radius of curvature). Load was applied on the prepared CeO$_{2}$/Pt/YSZ sample at a rate of 200 mm/s up to a peak load ($P_{max}$) of 9000 $\mu$N.

Figure S1 shows the load (\textit{p}) versus displacement (\textit{h}) as \textit{p--h} curves of CeO$_{2}$/Pt/YSZ and Pt/YSZ recorded during indentation with a Berkovich tip. Permanent displacement induced by the indentation was observed both in CeO$_{2}$/Pt/YSZ ($\approx$ 100 nm in depth, \textit{h}$_{p}^{(a)}$) and in Pt/YSZ ($\approx$ 80 nm in depth, \textit{h}$_{p}^{(b)}$), even after unloading. The indentation made during the loading-unloading process represents the work that is dissipated through the plastic deformation and is absorbed into a material (i.e., toughness of the material). This area is $3.933 \times 10^{-10}$ J for CeO$_{2}$/Pt/YSZ, denoting that the externally applied mechanical energy was consumed and absorbed into CeO$_{2}$/Pt/YSZ during plastic deformation. In addition, the \textit{p-h} curves of CeO$_{2}$/Pt/YSZ and Pt/YSZ exhibit smooth loading-unloading curves without any detectable pop--in events, which is likely due to the presence of Pt as a ductile metal and the thin CeO$_{2}$ layers.

\includegraphics[width=140mm]{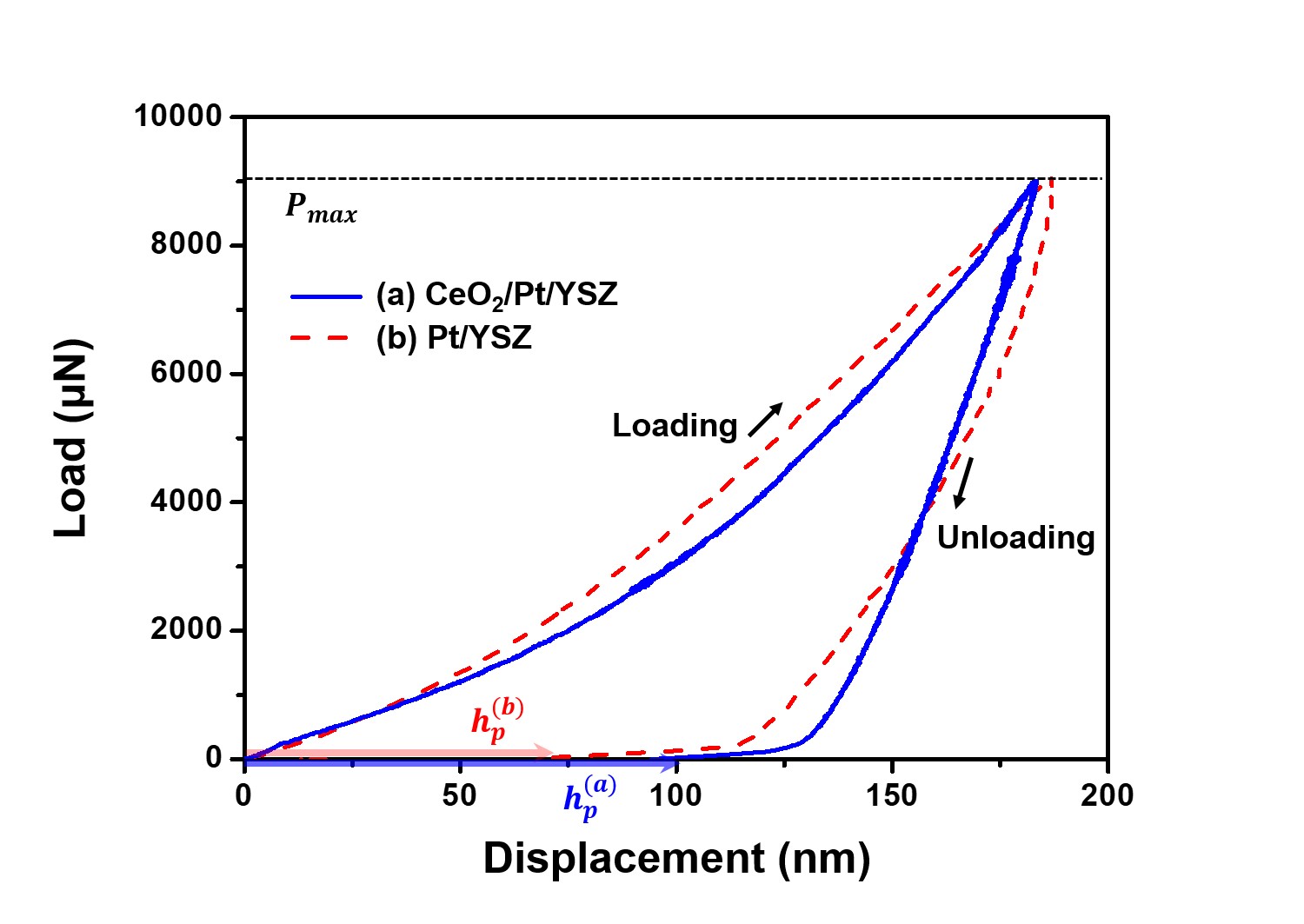}\\
\textbf{Figure S1} Load-displacement (\textit{p--h}) curves of \textbf{(a)} CeO$_{2}$/Pt/YSZ and \textbf{(b)} Pt/YSZ. \textit{h}$_{p}^{(a)}$ and \textit{h}$_{p}^{(b)}$ are permanent depth in CeO$_{2}$/Pt/YSZ and Pt/YSZ, created after removal of the test force. $P_{max}$ is the applied peak load.\\
\\
\\
\\
\large{\textbf{C--AFM Sampling}}\\
\\
Figure S2a shows a photograph of an indented CeO$_{2}$/Pt/YSZ sample attached on a sample puck assembly (part number: 448.140) using silver (Ag) paste for C--AFM. Figure S2b is a schematic figure showing the indented CeO$_{2}$/Pt/YSZ attached on a sample puck assembly (Figure S2a). To electrically bias the sample, the Ag paste (light gray color in Figure S2b) was also spread on the Pt film layer (dark gray color in Figure S2b), so that the Ag paste and the Pt layer beneath the CeO$_{2}$ thin film can be connected. Since the current from the tip goes through the CeO$_{2}$ and Pt layers (not to the YSZ substrate) in sequence, the contrast observed in the current image (obtained in the room temperature C--AFM analysis) is considered only from the electronic current, which is related to small polaron hopping.
\\
\\
\includegraphics[width=110mm]{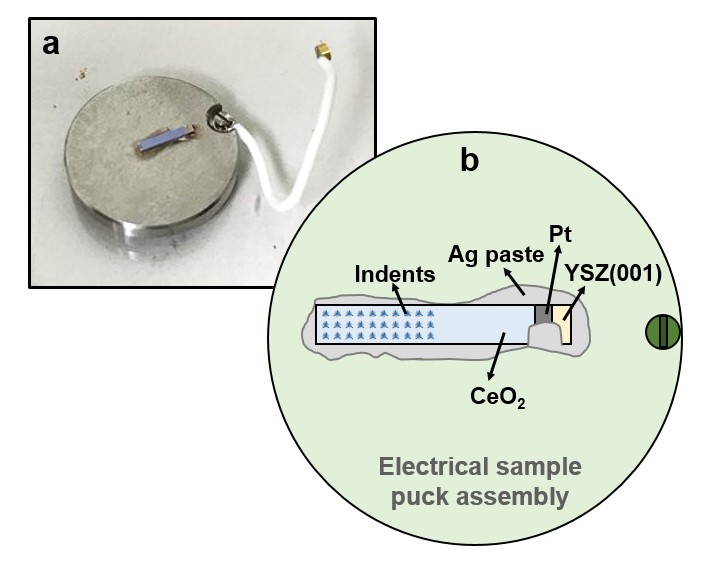}\\
\textbf{Figure S2 (a)} A schematic picture of an indented CeO$_{2}$/Pt/YSZ attached on a sample puck assembly by Ag paste for C-AFM analysis. \textbf{(b)} An explanatory figure showing the sample attached on the sample puck assembly, shown in more detail in figure (a).\\
\\
\\
\\
\large{\textbf{TEM sampling with SEM/FIB}}\\
\\
A TEM sample of indented CeO$_{2}$/Pt/YSZ was prepared using a focused ion beam (FIB, FEI Helios 600 NanoLab). Prior to placing the indented CeO$_{2}$/Pt/YSZ in the vacuum chamber of FIB/SEM, Pt/Pd (80:20; 30 nm thick) was deposited on the indented CeO$_{2}$/Pt/YSZ using a sputter coater (EMS 300TD) to eliminate the effect of charge and to protect the sample surface from Ga$^{+}$ ion beam damage. To extract the indented area in CeO$_{2}$/Pt/YSZ for preparing TEM samples, the indent point was first found in the SEM-FIB, as shown in Figure S3a. To protect the sample surface from the ion source, 500 nm thick Pt was deposited by electron beam, followed by the ion beam-mediated deposition of Pt (2 $\mu$m thick) using a gas injection system (GIS), as shown in Figure S3b. The sample was milled using the Ga$^{+}$ ion at 30 kV and various currents. A tungsten omniprobe was used to pick up the milled plate and to attach it on the copper half TEM grid (Ted Pella, Omniprobe Lift-Out grids) (Figure S3c). After sample attachment on the grid, fine milling was conducted at low currents by gradually reducing the milling current (Figure S3d). Afterwards, fine milling at reduced ion energy (Ar source, 500 eV) was carried out to remove the surface amorphous layer (Fischione NanoMill 1040).
\\
\\
\includegraphics[width=160mm]{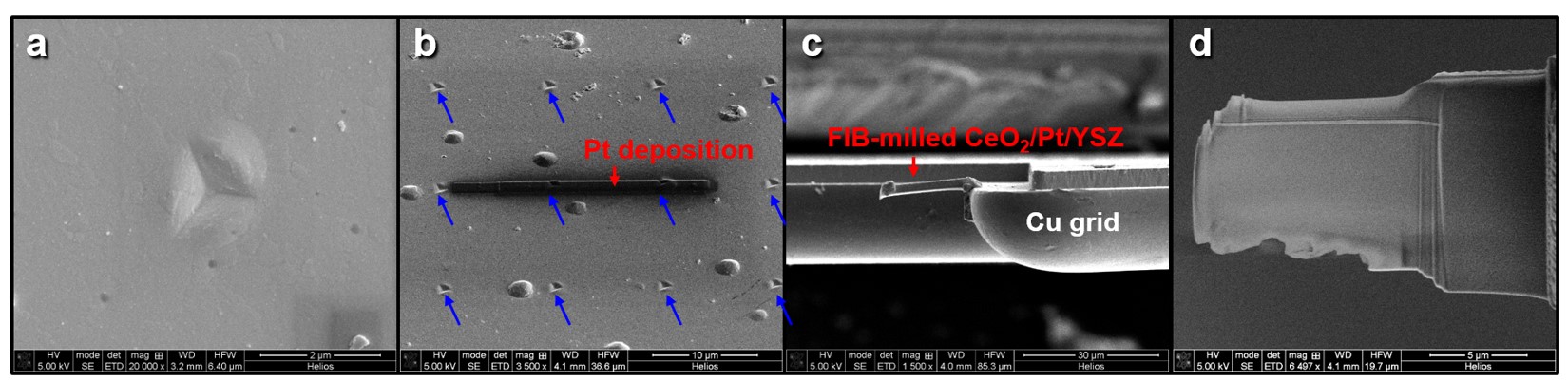}\\
\textbf{Figure S3} TEM sampling procedure of an indented CeO$_{2}$/Pt/YSZ with FIB/SEM. SEM images of \textbf{(a)} a surface of an indent on the CeO$_{2}$/Pt/YSZ, \textbf{(b)} the indented CeO$_{2}$ thin film after deposition of Pt, using GIS, \textbf{(c)} Ga$^{+}$ ion-milled CeO$_{2}$/Pt/YSZ plate attached on Cu half grid using tungsten omniprobe, and \textbf{(d)} TEM sample milled more from figure (c) at lower milling currents.\\
\\
\\
\\
\\
\large{\textbf{Selection of CeO$_{2}$ supercell size}}\\
\\
Figure S4a shows the stress-strain curves obtained during uniaxial compression of the CeO$_{2}$ supercell (aspect ratio c/a = 2) with classical MD simulations at 298 K with different supercell sizes. Periodic boundary conditions were applied in all directions for the supercells used for uniaxial compression. The deformation behaviors are very similar, independent of the supercell size, while the atomic configuration beyond the yield point ($\approx$ 8.5\%) differs depending on the size (Figure S4b); the smallest supercell consisting of 216 atoms is uniformly deformed and changed into a new phase (referred to as T-CeO$_{2}$) in the entire sample beyond the elastic limit. When the supercell is too small, the periodic boundary conditions cannot capture the long-range stress and strain fields that form in the realistic bulk sample. However, in supercells larger than the supercell with 216 atoms, i.e., the supercells consisting of 384, 600, and 2400 atoms (Figure S4b), inhomogeneous plastic deformation behaviors are captured (Figure S4c). Therefore, the supercell composed of 384 atoms is considered the minimum size for capturing both plastic deformation and its effect on the charge redistribution in CeO$_{2}$ under plastic deformation (Figures 5 in the main text and S14).
\\
\\
\includegraphics[width=160mm]{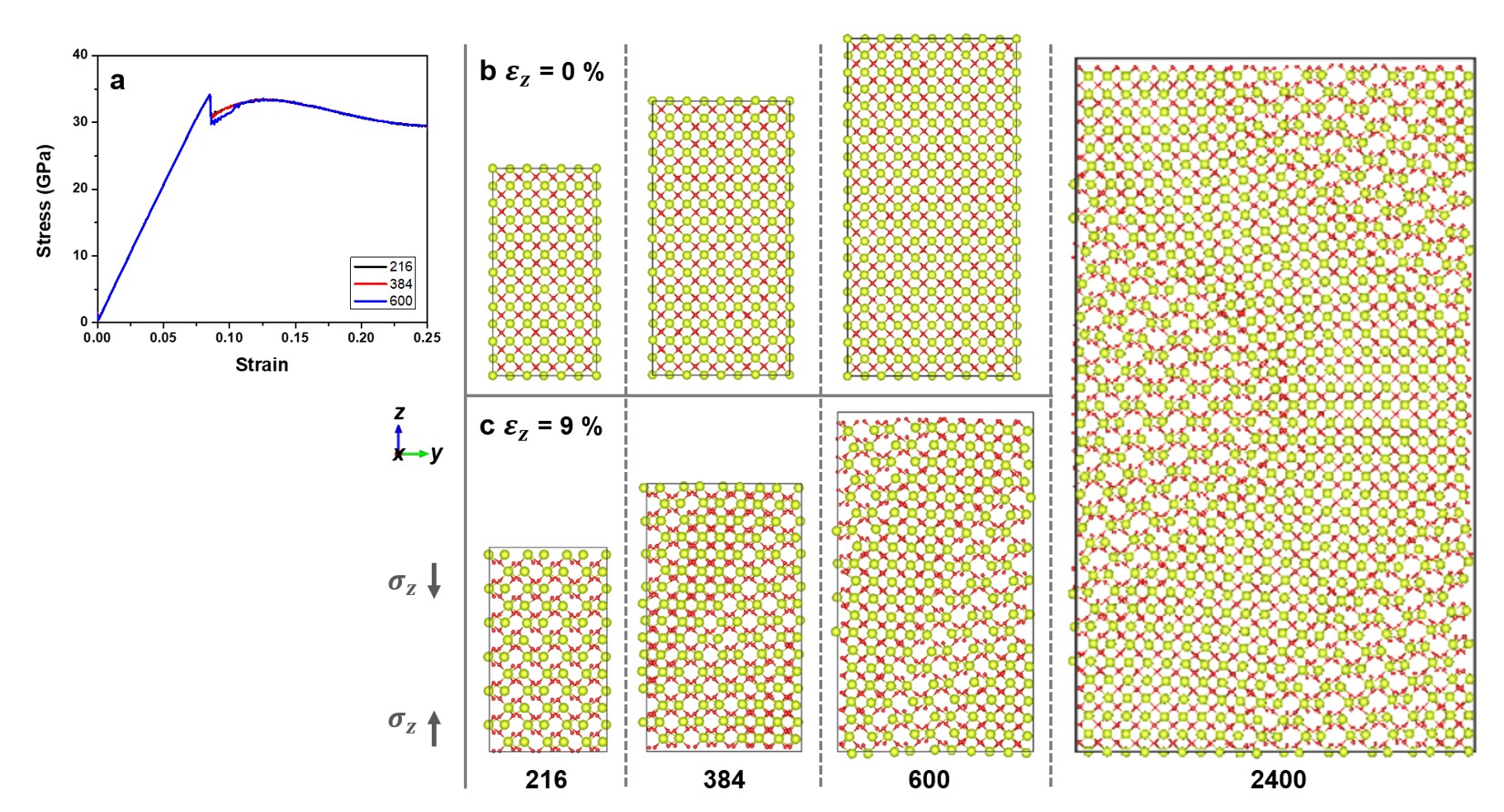}\\
\textbf{Figure S4 (a)} Stress-strain curves of different CeO$_{2}$ supercell sizes, obtained during uniaxial compression with MD simulations at 298 K. The number in the legend denotes the total number of atoms constituting the supercell (aspect ratio c/a = 2). (b) Initial structures of the supercells (consisting of 216, 384, and 600 atoms) (c) Atomic configurations of different sized CeO$_{2}$ supercells under the uniaxial compressive strain right beyond the elastic limit ($\epsilon_{z}$ $\approx$ 9\%). Yellow-green and red circles in figures (b) and (c) denote the Ce and O atoms, respectively.\\
\\
\\
\\
\\
\large{\textbf{Effect of a bias on current passing through a sample}}\\
\\
Figure S5 shows current images obtained from conductive tip--atomic force microscopy (C--AFM) analysis as a function of a bias applied to the indented CeO$_{2}$/Pt/YSZ sample with a Berkovich tip (Figure S1). With increasing magnitude of the applied bias, the current image has darker contrast on average, where darker contrast indicates a region with higher current passing through. Therefore, we confirm that electrical conductivity increases at the locally deformed area around the indent at room temperature (Figure S5). However, the increment is not exactly proportional to the applied bias, considering that the average current contrasts obtained from 3 V and 10 V are similar, even though the applied bias increases.
\\
\\
\includegraphics[width=160mm]{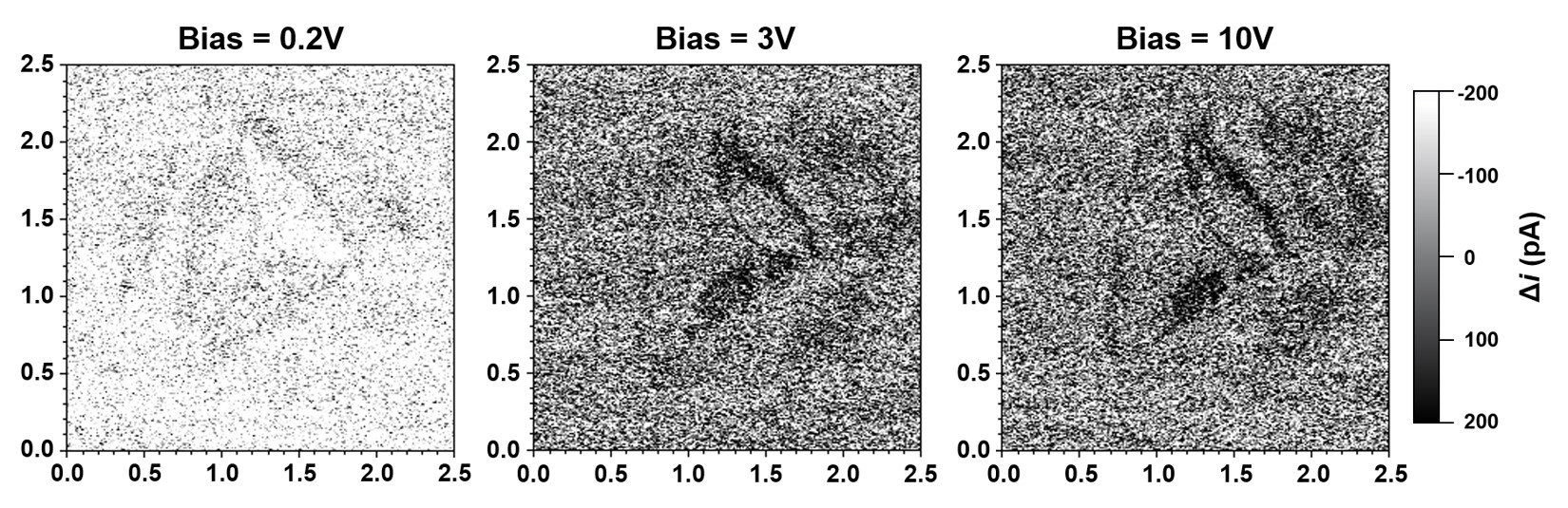}\\
\textbf{Figure S5} Current images obtained from C--AFM analysis at the applied bias of 0.2, 3, and 10 V. All current images were set at the same offset and current range, and the size of scanned images.\\
\\
\\
\\
\\
\large{\textbf{Pile--up vs. current map}}\\
\\
Figures S6a and b show deflection and current images of the indent on the indented CeO$_{2}$/Pt/YSZ obtained from C--AFM measured at 10 V. The indent generated by a Berkovich tip shows the pyramidal shaped pile--up (red dotted lines added as a visual guide) (Figure S6a). However, the current contrast observed around the pile-up region does not follow the indent and pile--up configuration (Figure S6b). This suggests that the contrast in the current map detected with C--AFM arises from a local deformation rather than the topological effect of the pile--up.
\\
\\
\includegraphics[width=130mm]{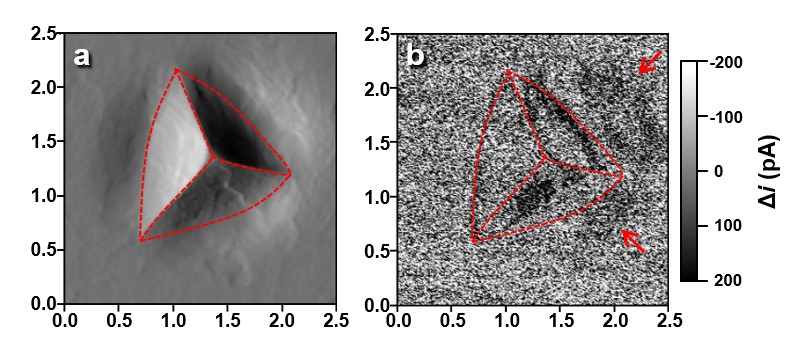}\\
\textbf{Figure S6 (a)} Deflection and \textbf{(b)} current images obtained from the C--AFM analysis. Red dotted lines show the pyramidal shape around the indent. The size of the scanned images is $2.5 \times 2.5$ $\mu$m$^{2}$.\\
\\
\\
\\
\\
\large{\textbf{Effect of indentation tip on current image}}\\
\\
Because significant topological effects can be introduced to Berkovich tip--indented samples, the indentation was performed on CeO$_{2}$/Pt/YSZ using a conical tip for comparison. The conical tip can uniaxially compress a material, limiting the indentation to a micrometer--sized local area, thus minimizing the potential topological effect introduced by the Berkovich tip. 
As can be seen in Figure S7a, the \textit{p-h} curve obtained from indentation with the conical tip (red dashed line) shows a permanent deformation similar to that obtained with the Berkovich tip (black line), after recovery of the large elastic deformation in the process of unloading. Uniform pile--up is generated around the indented point (Figure S7b). The current image does not exactly correspond to that of the height (topological) image, rather it forms circular current contrast around the permanently deformed indent point. In accordance with the current map of the Berkovich tip-indented CeO$_{2}$/Pt/YSZ (Figure 1 in the main text), the highest current contrast is observed near the pile--up around the indent point. However, the current contrast does not follow the pile--up aspect, indicating a negligible topographical effect on the current image around the indent. Hence, the current contrast observed in the indented CeO$_{2}$ layer originates from the locally varied electronic properties of the CeO$_{2}$ layer.
\\
\\
\includegraphics[width=160mm]{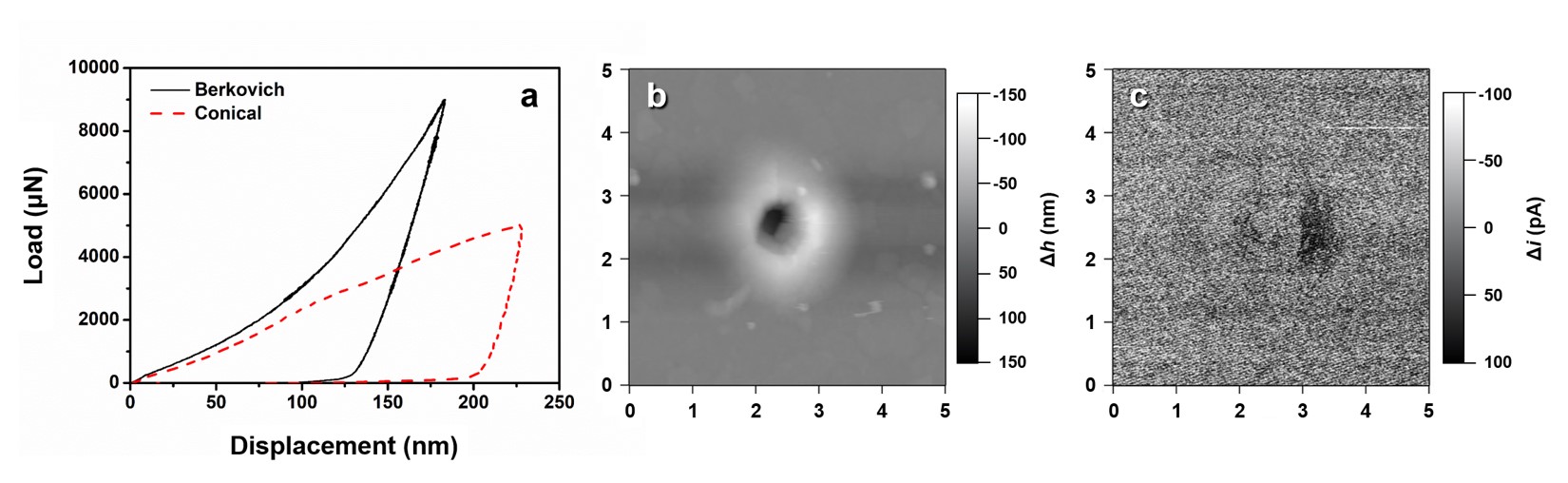}\\
\textbf{Figure S7 (a)} \textit{p-h} curve of CeO$_{2}$/Pt/YSZ indented with a conical tip. Load was applied on the prepared CeO$_{2}$/Pt/YSZ sample at a rate of 200 mm/s, up to peak load of 5000 $\mu$N. \textit{p-h} curve of the CeO$_{2}$/Pt/YSZ indented with a Berkovich tip is superimposed on figure (a) for comparison. \textbf{(b)} Height and \textbf{(c)} current images obtained from the C--AFM analysis of the conical tip--indented CeO$_{2}$/Pt/YSZ at a bias voltage of 3 V.\\
\\
\\
\\
\\
\large{\textbf{CeO$_{2}$/Nb:STO indentation and current contrast}}\\
\\
As observed in Figure 1 in the main text, the higher current around the indented area is also observed in another system, i.e., CeO$_{2}$ thin film (90 nm thick) deposited on Nb--doped SrTiO$_{3}$ substrate (Nb 0.7 wt\%:STO, MTI) indented with a Berkovich tip (Figures S8a--c). In accordance with Figure 1 in the main text, higher current contrast is observed in the locally deformed area around the indent, which directly confirms that the local deformation at room temperature causes the modification of electronic conductivity. This ensures that the inhomogeneous current contrast observed in indented CeO$_{2}$/Pt/YSZ is not an artifact.
\\
\\
\includegraphics[width=160mm]{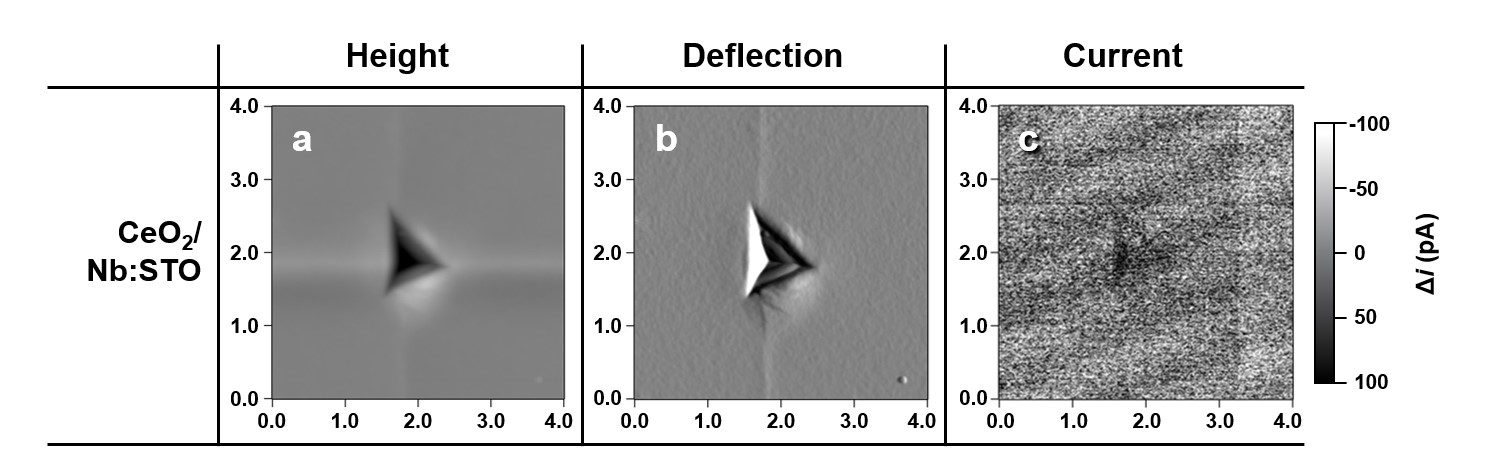}\\
\textbf{Figure S8 (a)} Height, \textbf{(b)} deflection, and \textbf{(c)} current images recorded by applying 10 V to the indented CeO$_{2}$/Nb:STO. The size of the scanned images is $4 \times 4$ $\mu$m$^{2}$.\\
\\
\\
\\
\large{\textbf{Polyhedral structures in T-CeO$_{2}$}}\\
\\
Figure S9 shows the atomic configurations of T-CeO$_{2}$, trigonal Ce$_{7}$O$_{12}$ (R$\overline{3}$), tetragonal CeO$_{2}$ (P4$_{2}$/mnm), cubic Ce$_{2}$O$_{3}$ (Ia$\overline{3}$), and cubic CeO$_{2}$ (Fm$\overline{3}$m) analyzed with polyhedra (shaded with sky blue faces). In the T-CeO$_{2}$ phase, 3--D octahedral short--range ordered (SRO) structures (marked with (i) in Figures S9a--h) are formed around Ce ions, with vertex and edge connections (sharing) between the neighboring octahedra (Figures S9a--b). These octahedral SRO structures with vertex and edge sharing are also observed in Ce$_{7}$O$_{12}$, tetragonal CeO$_{2}$ and cubic Ce$_{2}$O$_{3}$ (Figures S9c--h), but not in cubic CeO$_{2}$ (Figures S9i--j). Unlike the T-CeO$_{2}$, trigonal Ce$_{7}$O$_{12}$, tetragonal CeO$_{2}$, cubic Ce$_{2}$O$_{3}$ phase composed of both vertex and edge sharing among octahedral, cubic CeO$_{2}$ presents only edge sharing among hexahedra (marked with (ii) in Figures S9i--j). 
According to a variety of polyhedral structural factors summarized in Table S1, T-CeO$_{2}$ phase is similar to trigonal Ce$_{7}$O$_{12}$, tetragonal CeO$_{2}$ and cubic Ce$_{2}$O$_{3}$ among various Ce-O compounds, as marked in yellow in Table S1.
\\
\\
\includegraphics[width=160mm]{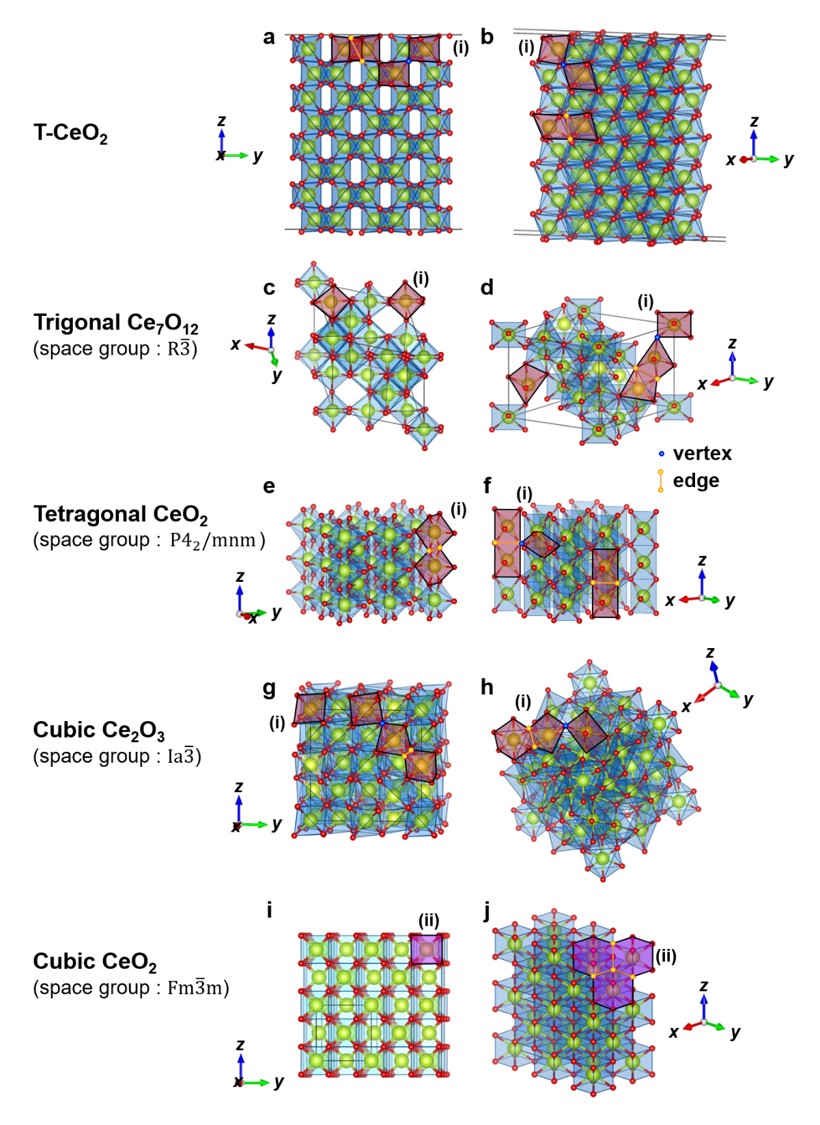}\\
\textbf{Figure S9} Atomic configurations of \textbf{(a)--(b)} T-CeO$_{2}$ obtained after plastic deformation ($\epsilon_{z}$= 8.5\% shown in Figure 4c in the main text), \textbf{(c)--(d)} trigonal Ce$_{7}$O$_{12}$ structure (space group : R$\overline{3}$), \textbf{(e)--(f)} tetragonal CeO$_{2}$ structure (space group : P4$_{2}$/mnm), \textbf{(g)--(h)} cubic Ce$_{2}$O$_{3}$ (space group : Ia$\overline{3}$) structure, and \textbf{(i)--(j)} cubic CeO$_{2}$ (space group : Fm$\overline{3}$m) structure along two different directions. The faces on the polyhedra found in T-CeO$_{2}$, Ce$_{7}$O$_{12}$, Ce$_{2}$O$_{3}$, and CeO$_{2}$ structures are shown with sky blue shade. The unit polyhedron observed in the structures are highlighted with red and purple shades as marked with (i) and (ii). The polyhedra shaded with red or purple are known as octahedron and hexahedron, respectively.\\
\\
\\
\textbf{Table S1} The polyhedral type constituting T-CeO$_{2}$, trigonal Ce$_{7}$O$_{12}$ structures, tetragonal CeO$_{2}$, cubic Ce$_{2}$O$_{3}$, cubic CeO$_{2}$, and its effective coordination number, volume, distortion index, and average Ce-O bond length in the polyhedra. The polyhedral structural factors close to that of the T-CeO$_{2}$ phase are shaded with yellow color.\\ 
\includegraphics[width=160mm]{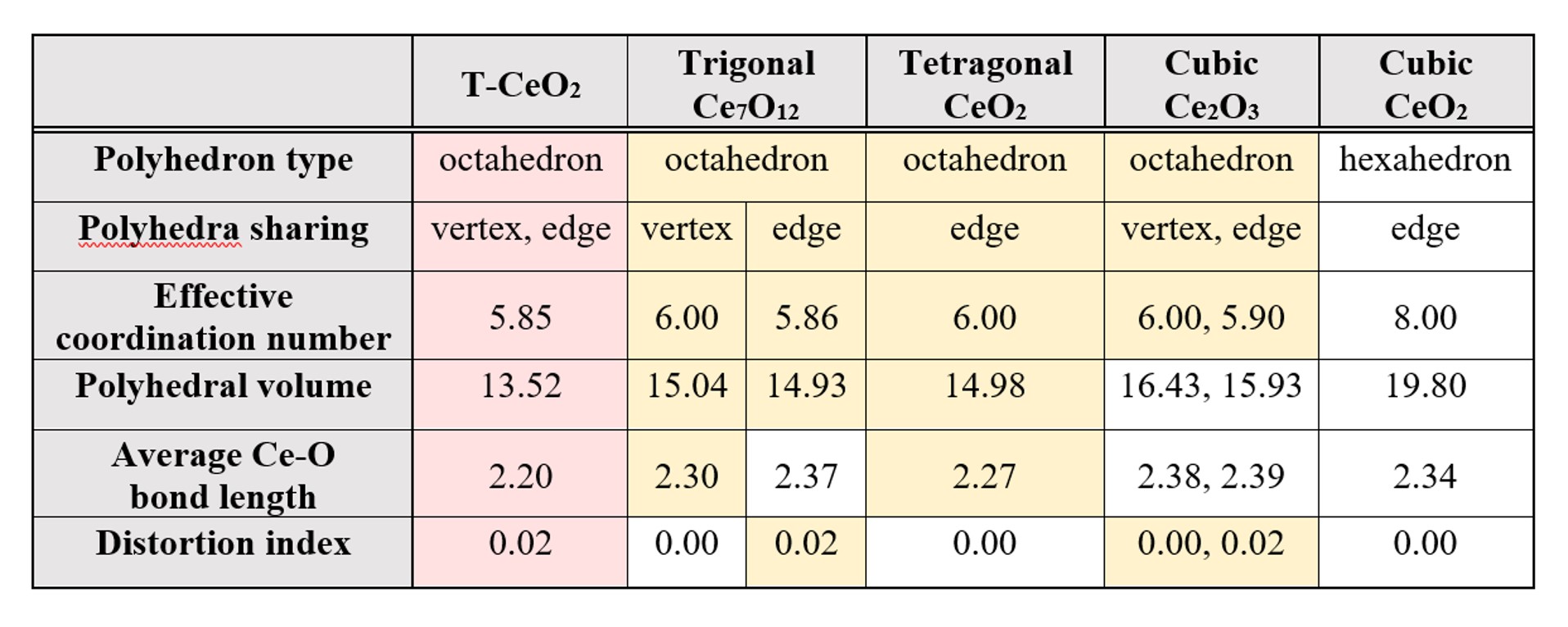}
$^{*}$The units of polyhedral volume and Ce-O bond length are \AA$^{3}$ and \AA. 
\\
\\
\\
\\
\large{\textbf{Presence of oxygen vacancy in CeO$_{2}$}}\\
\\
Figure S10a shows the atomic configurations of DFT--relaxed fluorite--structured CeO$_{2}$ ($2 \times 2 \times 2$ unit cells) containing one oxygen vacancy without straining. The relaxation method is the same as explained in the ‘Oxygen vacancy formation energy’ section in Methods in the main text. Around the oxygen vacancy (yellow sphere), several nearest neighbor oxygen ions (blue spheres) migrate slightly toward adjacent Ce ions or oxygen vacancy as marked with black arrows [1,2], making the Ce-O bond length shorter than in pristine CeO$_{2}$ (Table in Figure S10). Two Ce-O nearest neighbor bonds around any O ion (O5 in Figure S10a), especially those closest to oxygen vacancies among the four bonds, decrease the bond length slightly compared to that in pristine CeO$_{2}$ (2.403 $\rightarrow$ 2.358 \AA), while the other two Ce-O bonds slightly extend (2.403 $\rightarrow$ 2.529 \AA). This small change in the bond length between Ce-O, accompanied by the presence of oxygen vacancy, affects the electron redistribution around the oxygen vacancy (Figure S10b); excess electron density exists around the oxygen vacancy. It is well known that two electrons remain when an O ion leaves from the CeO$_{2}$ matrix, redistributed around two Ce ions near the O vacancy, changing the oxidation state of the Ce ion from 4+ to 3+ in order to satisfy the charge balance. Therefore, we expect that if the structure has bonding configurations (shorter or longer Ce-O bond lengths) similar to those near the oxygen vacancies in CeO$_{2}$ (Figure S10), then the inhomogeneous charge redistribution observed around the oxygen vacancy might occur in the structure (such as T-CeO$_{2}$ created during uniaxial plastic deformation, as shown in Figure 5 in the main text and Figure S14).
\\
\\
\includegraphics[width=140mm]{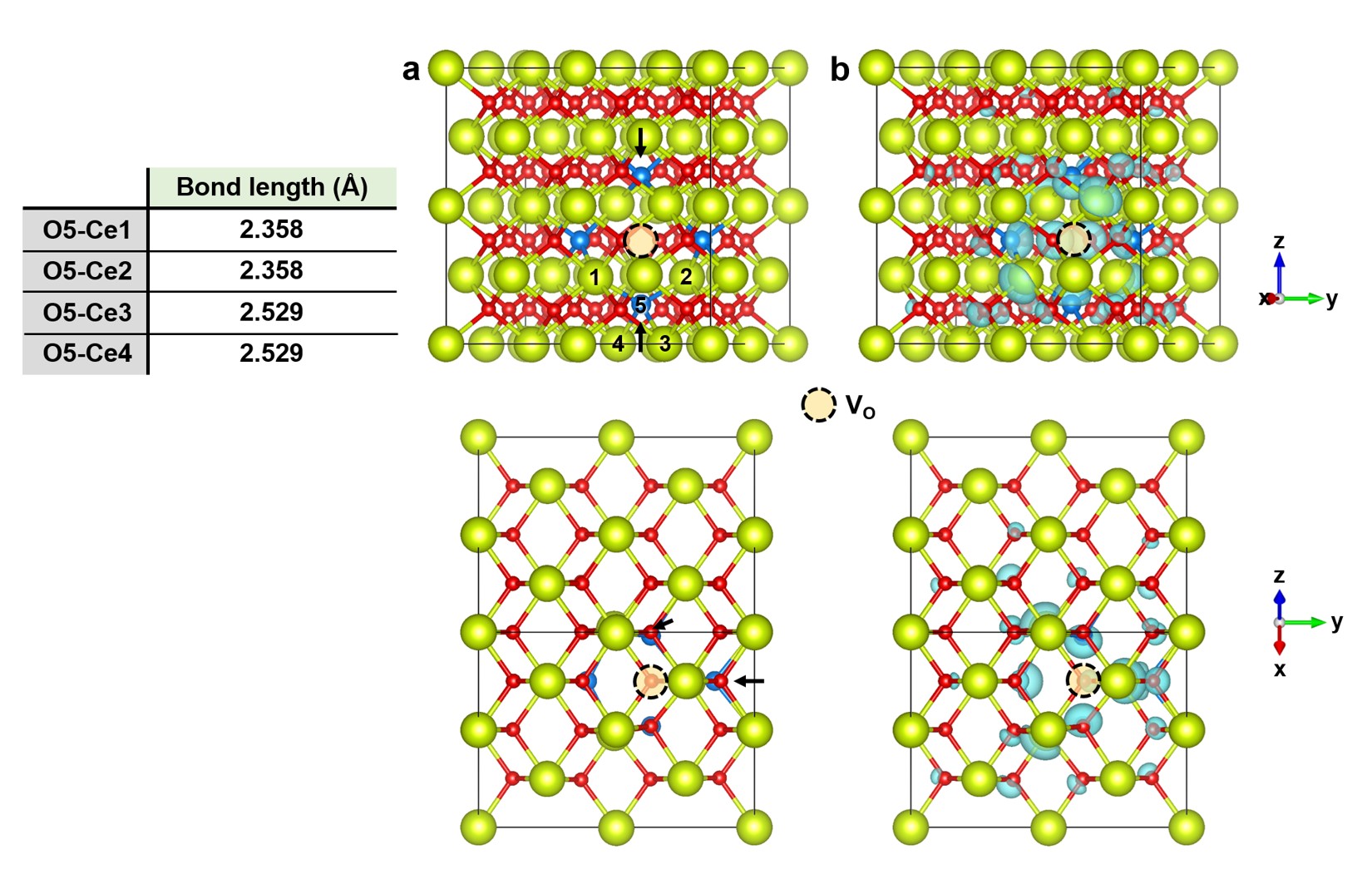}\\
\textbf{Figure S10 (a)} Atomic configuration and \textbf{(b)} differential charge distribution in fluorite structured CeO$_{2}$ consisting of $2 \times 2 \times 2$ unit cells (without straining) along $<$10 3 0$>$ (upper figures) and $<$1 0 1$>$ (bottom figures). The differential charge density ($\Delta\rho$) is defined as $\Delta\rho$=$\Delta\rho$(CeO$_{2}$+V$_{O}$)--$\Delta\rho$(CeO$_{2}$). Isosurface level (electron rich) was set at 0.03 e/\AA, with sky-blue color. Yellow-green and red circles are Ce and O atoms in CeO$_{2}$, while blue circles are O atoms which are nearest neighbors of the oxygen vacancy and move towards the oxygen vacancy. The yellowish circle marked with a dotted black line indicates the location of an oxygen vacancy in the CeO$_{2}$ structure.\\
\\
\\
\\
\large{\textbf{Change in Ce-O Bond length by uniaxial compression}}\\
\\
\textbf{Table S2} Bond length of Ce-O ions in a pristine CeO$_{2}$ structure, deformed CeO$_{2}$ phase in elastic and plastic regimes, and the T-CeO$_{2}$ phase in a plastic regime.
\\
\includegraphics[width=130mm]{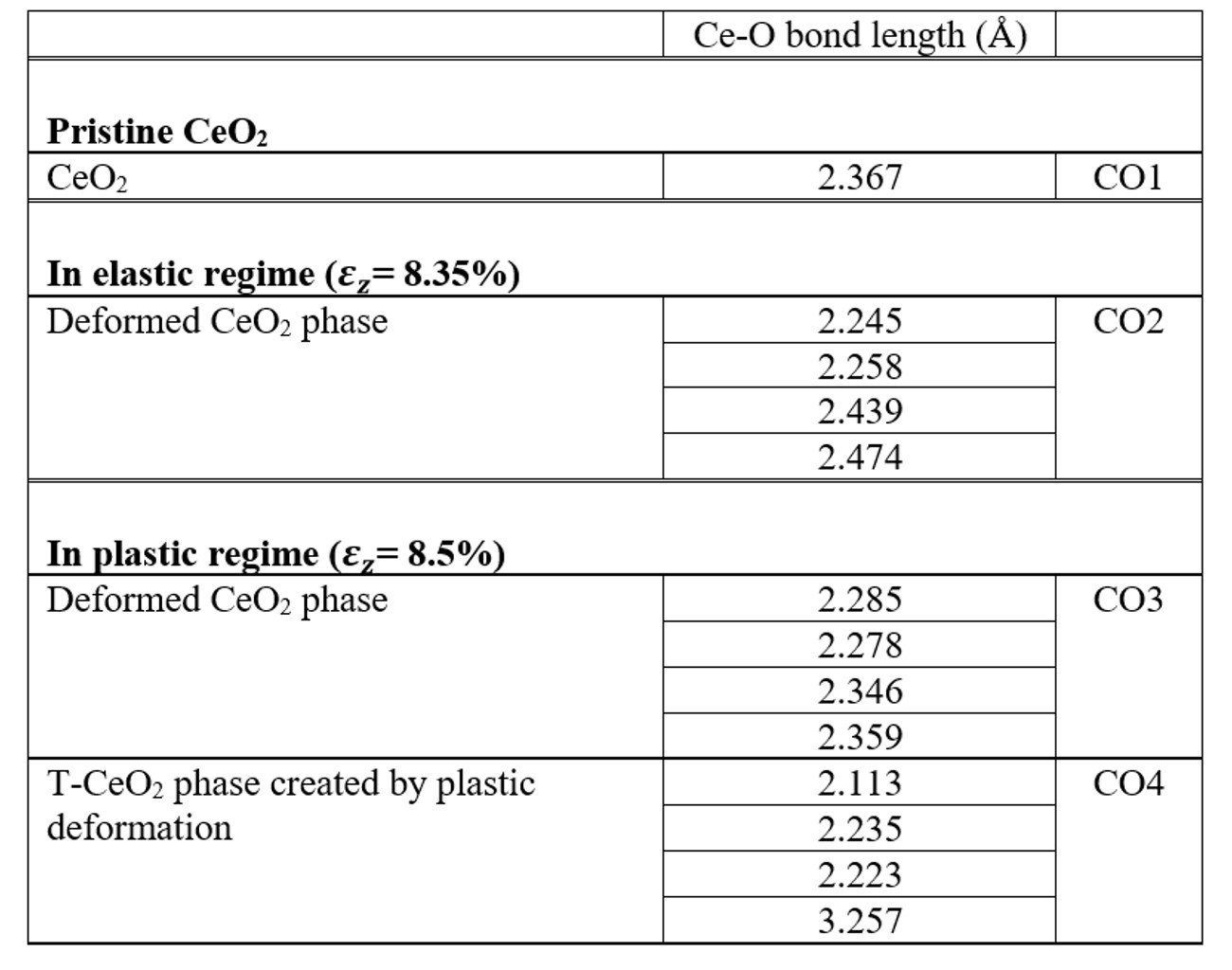}\\
\\
\\
\\
\large{\textbf{Change in the direction of CeO$_{2}$/T-CeO$_{2}$ interface}}\\
\\
Figure S11 shows snapshots of the CeO$_{2}$ supercell with c/a = 1 obtained during uniaxial compression, with respect to the applied compressive strain ($\epsilon_{z}$). A part of the CeO$_{2}$ supercell transforms into the T-CeO$_{2}$ phase (refer to the main text) beyond the elastic limit ($\epsilon_{z}$ = 8.42\%), with the interface of the CeO$_{2}$/T-CeO$_{2}$ phases parallel to the strain axis in the early stage of plastic deformation ($\epsilon_{z}$ = 8.45\%). However, the direction of the CeO$_{2}$/T-CeO$_{2}$ phase interface rotates 49 degree from the axis in a later stage ($\epsilon_{z}$ $\geq$ 8.55\%), as shown in Figure S11.
\\
\\
\includegraphics[width=160mm]{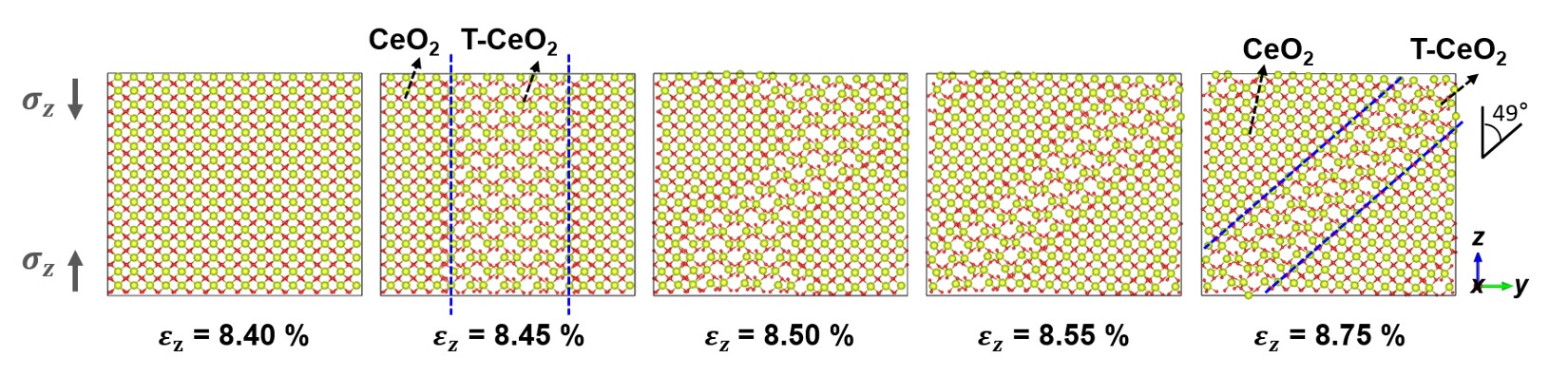}\\
\textbf{Figure S11} Movement of CeO$_{2}$/T-CeO$_{2}$ phase interface. Snapshots of CeO$_{2}$ supercell with c/a = 1, obtained during uniaxial compression, with respect to the applied compressive strain ($\epsilon_{z}$). Yellow-green and red circles in snapshots denote Ce and O atoms, respectively. The detailed explanation on the T-CeO$_{2}$ phase is in the main text.\\
\\
\\
\\
\large{\textbf{STEM analysis of CeO$_{2}$ thin film after indentation}}\\
\\
Figure S12 shows STEM-BF and STEM-HAADF images obtained around the indent point made on CeO$_{2}$ thin film (in the CeO$_{2}$/Pt/YSZ sample) at room temperature (marked with a red circle in the upper schematic figure in Figure S12). The STEM images are taken when the electron beam is aligned with the [110] zone axis of the CeO$_{2}$ thin film. The MD-simulated CeO$_{2}$ supercell in which the T-CeO$_{2}$ phase is generated during plastic deformation (refer to Figure 4c in the main text) is rotated along $<$110$>$ and shown in the right side of Figure S12a, to directly compare the STEM images to the MD simulated CeO$_{2}$ structure.

Unlike STEM images obtained from the undistorted CeO$_{2}$ thin film (far away from the indent, as marked in the upper schematic figure in Figure S13), several different contrasts are observed in the CeO$_{2}$ thin film deformed by indentation (Figure S12), as indicated with a, b, and c in the STEM images in Figure S12. Region a is the CeO$_{2}$ structure maintaining the initial cubic structure even though the thin film experiences plastic deformation upon indentation (Figure S12a). Ce ions in the CeO$_{2}$ phase (Figure S12a) exhibit diamond structures (blue solid lines serve as a guide; illustrated in Figures (1-1) and (1-2) in Figure S12). The yellow, green, and orange circles in Figure (1-2) are superimposed on Figure S12a in order to show that the material in the STEM images in Figure S12a has the same structure as pristine CeO$_{2}$. The STEM-BF contrast roughly estimated from the CeO$_{2}$ structure composed of heavy Ce ions and light O ions is shown in Figure (1-3) in Figure S12. Not only Ce ions, but also the contrast made by vacuum transmission or O ions (the brightest contrast in Figure (1-3)) exhibits the same diamond structure as indeed observed in the STEM-BF image in Figure S12a. Therefore, it is certain that region a is the CeO$_{2}$ structure oriented to the $<$110$>$ axis.

The diamond structure is also observed among Ce ions in the undistorted CeO$_{2}$ thin film (Figure S13). In addition, the bright contrast produced from O ions and vacant space (bright contrast in the BF image and dark contrast in the HAADF image) in the CeO$_{2}$ phase are also diamond shaped (Figure S13), confirming again that region a is a fluorite-structured CeO$_{2}$ phase.

Region c (Figure S12c) is the newly created T-CeO$_{2}$ phase that is transformed from the fluorite CeO$_{2}$ phase. Even though the CeO$_{2}$ phase has transformed into the T-CeO$_{2}$ phase under mechanical deformation, the Ce ions in the T-CeO$_{2}$ phase remain in the diamond structure both in the uniaxially compressed CeO$_{2}$ supercell with MD simulation (Figure 4d--e, right side of Figure S12a, (2-1) in Figure S12) and in the experimentally indented CeO$_{2}$ thin film (Figure S12c).

However, in region c, a different contrast is observed in the STEM-BF image in Figure S12c from that in Figure S12a; bright contrast in a zigzag pattern (not straight diamond symmetry) along the diagonal direction is observed in the STEM-BF image of the T-CeO$_{2}$ phase (same as the Figure (2-6) in Figure S12), which is expected to arise from the broken symmetry of O ions (including vacuum transmission) by movement of the O ions (illustrated in (2-2) to (2-5) in Figure S12). This zigzag contrast along the diagonal has never been observed for cubic CeO$_{2}$, and we did not observe any uneven (irregular) contrast in the undistorted region in the CeO$_{2}$ thin film (Figure S13).

The STEM contrast, especially in the BF image shown in region b, seems to have a strong strain field by deviating from the initial CeO$_{2}$ phase, but gradually changing from CeO$_{2}$ to T-CeO$_{2}$ while retaining the bonds between phases in regions a and c. The contrast observed in region b is frequently observed only in the indented CeO$_{2}$ thin film but not in the region far away from the indent point (Figure S13) when we randomly imaged around the indent point (all of the images are not shown in this manuscript). The region far away from the indent point (expected to be an undistorted region) only shows the CeO$_{2}$ structure in the STEM images (Figure S13).
\\
\\
\includegraphics[width=160mm]{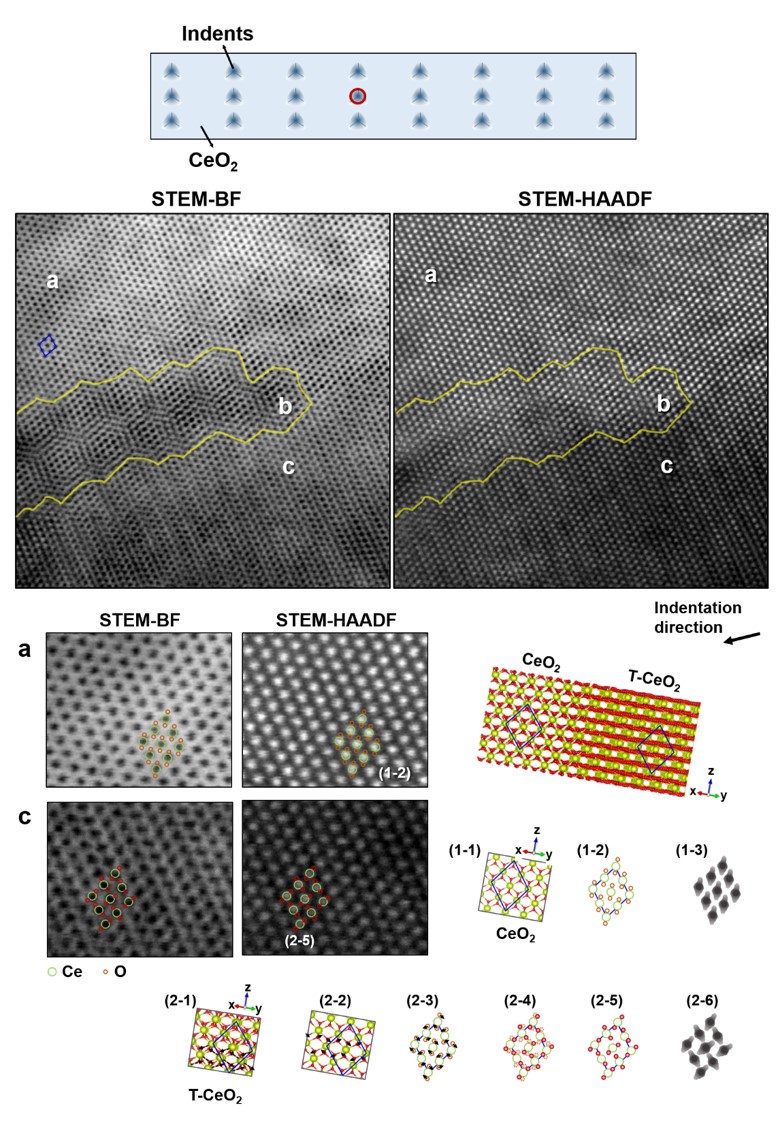}\\
\textbf{Figure S12} Upper figure : a schematic figure illustrating the location where the STEM images are taken from the indented CeO$_{2}$ thin film (in CeO$_{2}$/Pt/YSZ). The red open solid line denotes that the STEM images are taken from near the indent (deformed region) made on the CeO$_{2}$ thin film with a Berkovich tip. The TEM sampling and analysis methods are explained in Methods in the main text. Several different contrasts observed in the indented CeO$_{2}$ thin film are denoted as a, b, c in the STEM-BF and STEM-HAADF images. The magnified views (BF and HAADF images) of the regions a and c are shown in figures (a) and (c). The figure at the right side of figure (a) indicates the CeO$_{2}$ supercell uniaxially compressed with MD simulation (aspect ratio = 0.5 as shown in Figure 4b in the main text) but rotated along $<$110$>$. \textbf{(1-1)} Pristine CeO$_{2}$ structure aligned to $<$110$>$, but slightly compressed along the z-axis. \textbf{(1-2)} Schematic of atomic arrangement in figure (1-1). Yellow-green and orange open circles denote Ce and O ions located in figure (1-1). Figure (1-2) is superimposed on figure (a) (both in BF and HAADF images) to explain that the STEM images taken from the indented CeO$_{2}$ thin film match the CeO$_{2}$ structure shown in figure (1-2). \textbf{(1-3)} The STEM-BF image roughly predicted from the atomic arrangement shown in figure (1-2) for improved understanding of the STEM-BF image in figure (a). A heavier atom appears in darker contrast in the BF image. \textbf{(2-1)} The newly created T-CeO$_{2}$ structure during uniaxial compression with MD simulation shown in Figure 4h in the main text. \textbf{(2-2)} Schematic figure illustrating how O ions move (indicated with black arrows) in the T-CeO$_{2}$ phase, which is predicted by comparing the T-CeO$_{2}$ region in figure (2-1) with the pristine CeO$_{2}$ structure in figure (1-1). \textbf{(2-3)} Simplified figure of figure (2-2). \textbf{(2-4)} and \textbf{(2-5)} Schematics of the atomic arrangement in the final T-CeO$_{2}$ structure, estimated from the O ions movement illustrated in figure (2-3). The solid red open circles are the final location of O ions, while the orange dotted open circles are the initial position of O ions before phase transformation. Figure (2-5) is superimposed on figure (c) (both in BF and HAADF images) to explain that the STEM images experimentally taken match the CeO$_{2}$ structure shown in figure (2-5). \textbf{(2-6)} STEM-BF image roughly predicted from the atomic arrangement shown in figure (2-5) for improved understanding of the STEM-BF image in figure (c).\\
\\
\\
\includegraphics[width=160mm]{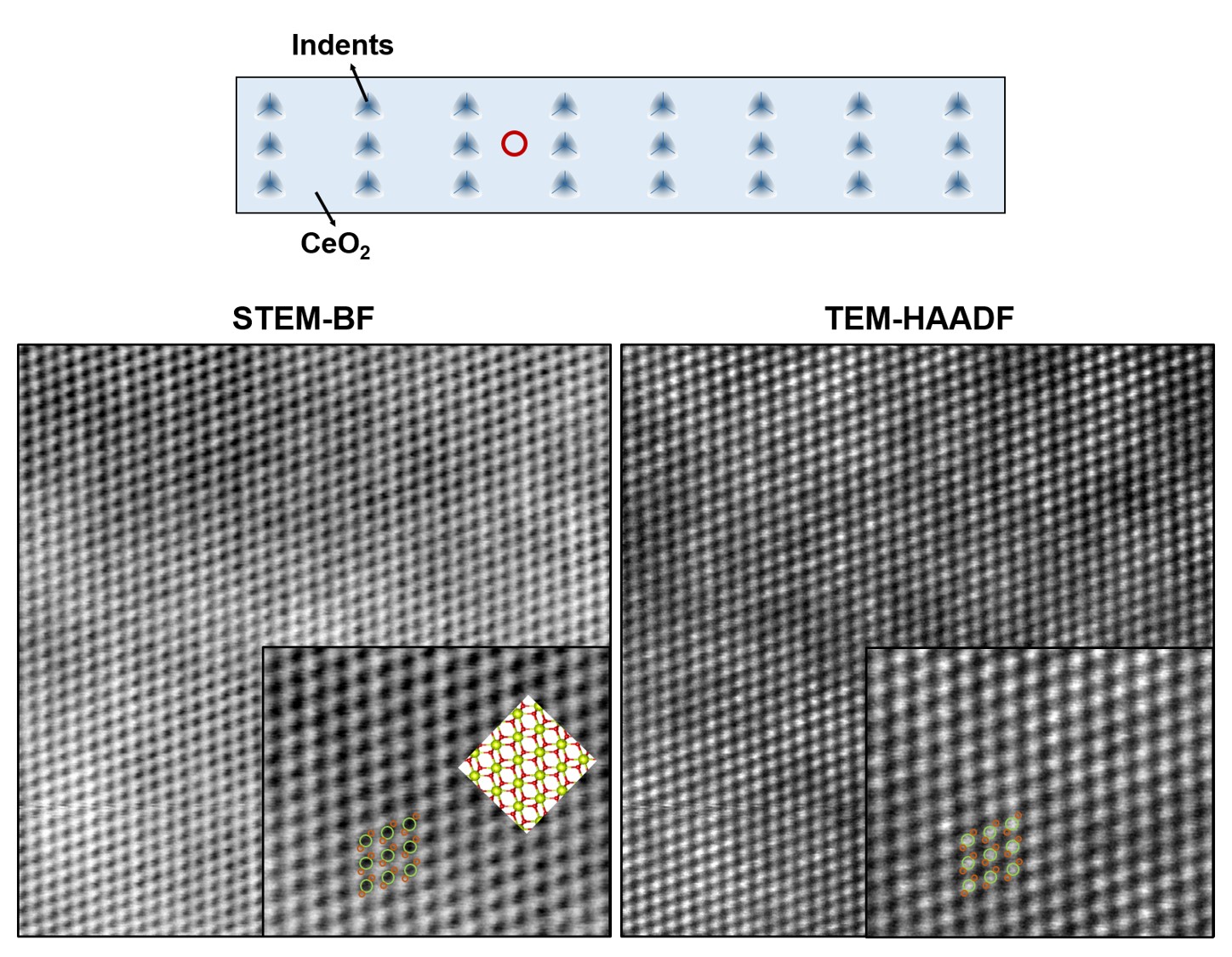}\\
\textbf{Figure S13} Upper figure: a schematic figure indicating the location where the STEM images are taken. A red open solid line denotes that the STEM sample is taken from the undistorted CeO$_{2}$ thin film (far away from the indents and therefore was not affected by the indentation). STEM-BF and STEM–HAADF images of the pristine CeO$_{2}$ structure and insets of their magnified views. Schematic of atomic arrangement in figure (1-2) (yellow-green and orange open circles) is superimposed to explain that the STEM images taken experimentally match the CeO$_{2}$ structure shown in figure (1-2).\\
\\
\\
\\
\large{\textbf{Charge redistribution in CeO$_{2}$ supercell with high aspect ratio}}\\
\\
Figure S14s shows the charge difference ($\Delta\rho$), i.e., the change in atomic charge compared to the average charge of each snapshot, defined as $\Delta\rho$(\textit{i},$\epsilon$) = $\rho$(\textit{i},$\epsilon$) – $\overline{\rho}$($\epsilon$), where the total charge of atom \textit{i} was integrated in a constant radius (volume) from the core. Static DFT calculations were conducted to investigate the atomic charge of individual atoms (\textit{i}) in the uniaxially compressed supercell (Figures S14a--d) with respect to strain using the same parameters as DFT calculations performed for the bulk CeO$_{2}$ supercell explained in the Methods in the main text. 
Within the elastic regime (up to $\epsilon_{z}$ = 8.6\%, Figures S14a--c), there is no large change in atomic charge in the entire supercell, whereas it suddenly exhibits a large difference right beyond the elastic limit in Figure S14d ($\epsilon_{z}$ = 9\%). The charge difference in the plastically deformed CeO$_{2}$ (T-CeO$_{2}$ in Figure S14e) is inhomogeneous, such that in the local area in which more severe plastic deformation occurs (T-CeO$_{2}$ phase), Ce ions tend to receive more electrons (displayed with more red color). The increased charge in the T-CeO$_{2}$ phase seems to arise from the less deformed region (CeO$_{2}$ region), considering that the CeO$_{2}$ phase loses atomic charges (blue color) (Figure S14d). This indicates that the deformation-induced structural change from CeO$_{2}$ to T-CeO$_{2}$ (similar to the formation of V$_{O}^{2+}$ in CeO$_{2}$ in the structural point of view) causes the redistribution of electrons such that electrons migrate from CeO$_{2}$ to T-CeO$_{2}$.
The CeO$_{2}$/T-CeO$_{2}$ interface is 5 -- 45 degree with respect to the uniaxial compression axis (z-axis), which differs from the CeO$_{2}$/T-CeO$_{2}$ interface parallel to the compressive axis observed in a supercell with a lower aspect ratio (c/a = 0.5) in Figures 4--5 in the main text.
\\
\\
\includegraphics[width=160mm]{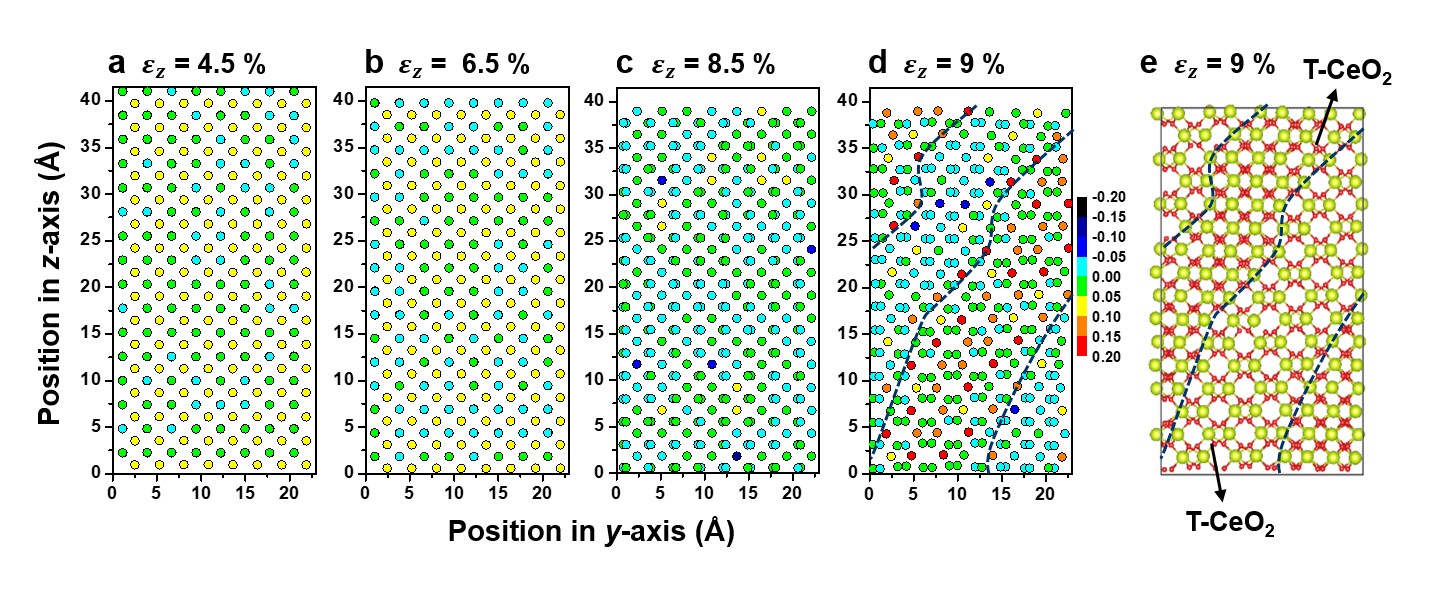}\\
\textbf{Figure S14} The distribution of charge difference in CeO$_{2}$ (composed of 384 atoms, c/a = 2) with respect to the applied uniaxial compressive strain, along the z-axis; \textbf{(a)} 4.5\%, \textbf{(b)} 6.5\%, \textbf{(c)} 8.5\% and \textbf{(d)} 9\%. Here, one circle denotes the position of an atom. The color map (right side of figure (d)) is identical for figures (a)--(d), and the unit is e/atom. \textbf{(e)} Atomic configuration of the deformed CeO$_{2}$, at $\epsilon_{z}$ = 9\%. Yellow-green and red circles in figure (e) denote Ce and O atoms, respectively. The elastic limit in the CeO$_{2}$ supercell with c/a = 2 is $\epsilon_{z}$ = 8.6\%. The black dotted lines in figures (d) and (e) are the CeO$_{2}$/T-CeO$_{2}$ phase interfaces.\\
\\
\\
\large{\textbf{Supplementary references}}\par 
1. Z. Yang, G. Luo, Z. Lu and K. Hermansson, J. Chem. Phys., 2007, 127, 074704.\\
2. M. Alaydrus, M. Sakaue and H. Kasai, Phys. Chem. Chem. Phys., 2016, 18, 12938--12946.
\end{document}